\documentclass[twocolumn,pra,showpacs,superscriptaddress]{revtex4}
\usepackage{amssymb}
\usepackage{amsmath}
\usepackage{graphicx}
\usepackage{subfigure}
\usepackage{natbib}
\usepackage{epsfig}
\usepackage{amsfonts}
\usepackage{mathrsfs}
\usepackage{ulem}
\usepackage{color}
\usepackage[toc,page,title,titletoc,header]{appendix}

\begin{document}

\title{Scattering and effective interactions of ultracold atoms with spin-orbit coupling}
\author{Long Zhang}
\affiliation{Hefei National Laboratory for Physical Sciences at Microscale and Department
of Modern Physics, University of Science and Technology of China, Hefei,
Anhui 230026, China}
\affiliation{Department of Physics, Renmin University of China, Beijing, 100190, China}
\author{Youjin Deng}
\affiliation{Hefei National Laboratory for Physical Sciences at Microscale and Department
of Modern Physics, University of Science and Technology of China, Hefei,
Anhui 230026, China}
\author{Peng Zhang}
\email{pengzhang@ruc.edu.cn}
\affiliation{Department of Physics, Renmin University of China, Beijing, 100190, China}

\begin{abstract}
We derive an analytical expression for the scattering amplitude of two
ultracold atoms of arbitrary spin and with general spin-orbit (SO) coupling,
on the basis of our recent work (Phys. Rev. A \textbf{86}, 053608 (2012)).
As an application, we demonstrate that SO coupling can induce scattering
resonance in the case with finite scattering length. The same approach can
be applied to calculate the two-body bound state of SO-coupled ultracold
atoms. For the ultracold spin-$1/2$ fermi gases in three- or two- dimensional
systems with SO coupling, we also obtain the renormalization
relation of effective contact interaction with momentum cutoff, as well as
the applicability of Huang-Yang pseudo-potential.
\end{abstract}

\pacs{03.65.Nk, 34.50.-s, 05.30.Fk}
\maketitle

\section{Introduction}

In the study of ultracold gases it is very important to understand the
low-energy scattering properties of atoms. First, the two-body and
three-body collisions are the underlying physics of many important
experimental phenomena, e.g., the two-body decay and three-body
recombination. Second, understanding the behavior of the low-energy
inter-atomic scattering amplitudes is indispensable in designing the
effective inter-atomic interactions (e.g., the Huang-Yang pseudo-potential~%
\cite{hyp}, Bethe-Peierls boundary condition~\cite{bpc} and the contact
interaction with momentum cutoff~\cite{pr}), which are widely used in the
theoretical calculations. In the low-energy cases the effective interaction
and the realistic interaction potential should lead to the same two-body
scattering amplitude.

In recent years, a class of synthetic gauge fields and spin-orbit (SO)
coupling has been realized in ultracold Bose gases~\cite%
{NIST,NIST_elec,SOC,collective_SOC,NIST_partial,JingPRA,ourdecay} and
degenerate Fermi gases~\cite{SOC_Fermi,SOC_MIT,jing13} with Raman laser beams~\cite%
{sth,liu}. In these systems, the atomic spin is linearly coupled with
the aotmic spatial momentum. A considerable amount of theoretical interest
has been stimulated to understand the SO-coupling effect in both many-body~%
\cite{review,Stripe,Ho,Wu,Victor,Hu,Santos,vyasanakere-11, gong-11, yu-11,
hu-11, iskin-11, yi-11, han-12, dellanna-11, zhou-11b, chen-12, seo-11,
huang-11, he-11,m-2,zhang-12,hu13a,hu13b,hu13c,hu13d,hu13e,hu13f,fewbody0} and few-body physics~\cite%
{fewbody1,fewbody2,fewbody3,fewbody4,fewbody5,fewbody6,fewbody7,fewbody8,
fewbody0,ourbp,duan, fewbody9}%
.

In this paper we provide a systematic investigation on the two-body
scattering amplitude of SO-coupled ultracold atoms in three-dimensional (3D)
uniform space. Our research is based on the following motivations. First, in
the current experiments, the amplitude of elastic interatomic collision \cite%
{NIST_partial} and inelastic-scattering-induced decay \cite%
{NIST_partial,ourdecay} of ultracold gases have been directly observed in
the systems with SO coupling. The theoretical investigation for the two-body
scattering amplitude is necessary to explain this kind of observations. In
particular, the calculation of the inelastic scattering amplitudes is
crucial for the study of stability of the SO-coupled ultracold gases in the
metastable dressed state, e.g., the dark state. Second, as shown above, the
criteria of the effective interaction in the ultracold gases is that, the
effective interaction potential and the real interaction potential should
lead to the same low-energy two-body scattering amplitude. Accordingly, we
should first calculate the scattering amplitude given by the real potential,
and then construct the correct effective interactions. This kind of work has
been done for ultracold gases in quasi-one-dimensional~\cite{osc},
quasi-two-dimensional~\cite{petrov01} confinements and optical lattices~\cite{o1,o2}, but it
is still absent  for the gases with SO coupling.

The calculation in this paper is based on our recent work \cite{ourbp} where
the short-range behavior of the scattering wave functions of two SO-coupled
ultracold atoms in a 3D uniform system is studied and a modified
Bethe-Peierls boundary condition is derived. Based on these results, in this
paper we derive an analytical expression for the scattering amplitude of two
atoms with arbitrary spin and SO coupling. Our approach can also be used to
calculate the low-energy bound state of two spin-$1/2$ atoms with SO
coupling. Furthermore, we show that the SO coupling can induce the
scattering resonance. Namely, for the atoms with finite scattering length,
the threshold scattering amplitude diverges when the SO-coupling intensity
assumes some particular value. For the SO-coupled spin-$1/2$ fermonic atoms,
we also derive the renormalization relation of 3D and pure-two-dimensional
(pure-2D) effective contact interaction with momentum cutoff, as well as the
applicability of Huang-Yang pseudo-potential. We find that the form of the
3D and 2D renormalization relation is not changed by the SO coupling.
Nevertheless, in the presence of SO coupling the physical
parameters, i.e., the scattering length in the 3D case and bound-state
binding energy in 2D case, should be replaced by the ones which are related
to the SO coupling. Furthermore, we also find that the Huang-Yang
pseudo-potential cannot be directly used in the presence of SO-coupling.

The remainder of this manuscript is organized as follows. In Sec. II, we
derive the exact analytical expression for the scattering amplitude of
SO-coupled ultracold atoms, and show the approach for the calculation of
two-atom bound states. In Sec. III, we illustrate the scattering resonance
induced by SO coupling. The renormalization relation of effective contact
interaction and the applicability of Huang-Yang pseudo-potential is
investigated in Sec. IV. The main results are summarized and discussed in
Sec. V, while some details of our calculations are explained in the
appendixes.

\section{scattering amplitude and bound state of SO-coupled atoms}

\subsection{Spin-$1/2$ fermionic atoms}

We first consider the scattering amplitude of two SO-coupled spin-$1/2$
fermonic atoms in 3D space. In this paper we use the word
\textquotedblleft SO coupling" to refer to the linear coupling between
atomic spin and momentum, e.g., the ones realized in the current experiments
with Raman laser beams. Without loss of generality, the single-atom
Hamiltonian of such a system can be written as
\begin{equation}
H_{\mathrm{1b}}=\frac{\vec{P}^{2}}{2}+\lambda \vec{M}\cdot \vec{P}+Z\,,
\label{h1b}
\end{equation}%
where $\vec{P}$ is the atomic momentum, and $\vec{M}$ and $Z$ are operators
in spin space (here we have used $\hslash =1$ and the atomic mass $m=1$).
The term $\lambda \vec{M}\cdot \vec{P}$ describes the SO coupling and $Z$
accounts for the residual spin-dependent part. Here, the eigen-values of $%
\vec{M}$ are of the order of unity, and $\lambda $ indicates the intensity
of the SO coupling. It is pointed out that the term  $\lambda \vec{M%
}\cdot \vec{P}$ in Hamiltonian in Eq. (\ref{h1b}) can be used to
describe \textit{arbitrary} type of linear spin-momentum coupling. For
instance, for effective spin- $1/2$ systems in Refs.~\cite%
{SOC_Fermi,SOC_MIT}, one has  $\lambda =2k_{r}$\textbf{, }$\vec{M}=(\hat{%
\sigma}_{z},0,0)$ and $Z=\delta \hat{\sigma}_{z}/2+\Omega \hat{%
\sigma}_{x}/2$ where $\delta $ is the two-photon
detuning, $k_{r}$ is the recoil momentum, $\Omega $ is
the Raman-coupling strength, and $\hat{\sigma}_{x,y,z}$ is the
Pauli operator. Similarly, for the system with Rashba SO-coupling in the 3D
space, we have $\vec{M}=(\hat{\sigma}_{x},\hat{\sigma}_{y},0)$.

For the two-atom scattering problem, the Hilbert space $\mathcal{H}$ can be
expressed as $\mathcal{H}=\mathcal{H}_{r}\otimes \mathcal{H}_{s1}\otimes
\mathcal{H}_{s2}$, with $\mathcal{H}_{r}$ for the inter-atomic relative
motion in the spatial space, and $\mathcal{H}_{si}$\ ($i=1,2$) for the spin
of the $i$-th atom. In this paper we use $|\rangle \rangle $ to denote the
state in $\mathcal{H}$, $|)$\ for the state in $\mathcal{H}_{r}$,\ $|\rangle
$\ for the state in $\mathcal{H}_{s1}\otimes \mathcal{H}_{s2}$, and $%
|\rangle _{i}$\ for the state in ${\cal H}_{si}$. In Secs. II, III and Sec. IV.C we
work in the representation of inter-atomic relative position, where the
state $|\psi \rangle \rangle $\ is described by the \textquotedblleft spinor
wave function" $|\psi (\vec{r})\rangle \equiv (\vec{r}|\psi \rangle \rangle $%
. Here $|\vec{r})$\ is the eigen-state of the inter-atomic relative
position, with corresponding eigen-value $\vec{r}=(x,y,z)$. It is clear that
$|\psi (\vec{r})\rangle $\ can also be considered as a $\vec{r}$-dependent
spin state.

The total Hamiltonian of the two atoms is given by $H_{\mathrm{1b}}^{(1)}+H_{%
\mathrm{1b}}^{(2)}+U(\vec{r})$, where $H_{\mathrm{1b}}^{(i)}$ ($i=1,2$) is
for the $i$th atom, and $U(\vec{r})$ is the spin-dependent interaction
potential between the two atoms. Because the total momentum of the two atoms
is conserved, the relative motion of these two can be separated from their
mass-center motion. The Hamiltonian for the relative motion is then%
\begin{equation}
H=\vec{p}^{2}+\lambda \vec{c}{\cdot }\vec{p}+B(\vec{K})+U(\vec{r})\equiv
H_{0}+U(\vec{r}),  \label{hf}
\end{equation}%
where $\vec{p}$ is the relative-momentum operator of the two atoms. In the $%
\vec{r}$-representation we have $\vec{p}=-i\nabla $. The total momentum $%
\vec{K}=\vec{P}^{(1)}+\vec{P}^{(2)}$ of the two atoms is conserved during
the scattering process, and behaves as a constant in our calculation. The
operators $\vec{c}$ and $B(\vec{K})$ read
\begin{eqnarray}
\vec{c} &=&\vec{M}^{(1)}-\vec{M}^{(2)},  \label{c} \\
B(\vec{K}) &=&Z^{(1)}+Z^{(2)}+\frac{\lambda }{2}\vec{K}\cdot ({\vec{M}}%
^{(1)}+{\vec{M}}^{(2)})\,.
\end{eqnarray}

In the stationary scattering theory, the incident state is regarded
as the eigen-state of the Hamiltonian $H_{0}$ for the free motion of the two
fermionic atoms. With the Pauli's principle being taken into account, such
an incident state for two spin$-1/2$ fermonic atoms can be expressed as
\begin{equation}
|\Psi _{t}^{\left( 0\right) }(\vec{r})\rangle \!=\!\frac{e^{i\vec{k}\cdot
\vec{r}}}{4\pi ^{3/2}}|\alpha ,\vec{k}\rangle -\!\frac{e^{-i\vec{k}\cdot
\vec{r}}}{4\pi ^{3/2}}\mathrm{P}_{12}|\alpha ,\vec{k}\rangle \,,
\label{psi0}
\end{equation}%
with $\vec{k}$ the relative momentum of the two atoms and $\mathrm{P}_{12}$
the permutation operator of the spin of the two atoms. Here the state $%
|\alpha ,\vec{k}\rangle $ ($\alpha =1,2,3,4$) of the two-atom spin is
defined as the $\alpha $th eigen-state of the operator $h_{0}(\vec{k})\equiv
\lambda \vec{c}{\cdot }\vec{k}+B(\vec{K})$, with eigen-energy $\mathcal{E}%
(\alpha ,\vec{K},\vec{k})$. The symbol satisfies $\alpha \geq \alpha
^{\prime }$ when $\mathcal{E}(\alpha ,\vec{K},\vec{k})\geq \mathcal{E}%
(\alpha ^{\prime },\vec{K},\vec{k})$. In this paper, we denote
\begin{equation}
t=(\alpha ,\vec{K},\vec{k})
\end{equation}%
as the set of these three quantum numbers. It is easy to prove that $|\Psi
_{t}(\vec{r})\rangle \!$ in Eq.~(\ref{psi0}) is an eigen-state of $H_{0}$
with eigen-energy
\begin{equation}
E_{t}=k^{2}+\mathcal{E}(\alpha ,\vec{K},\vec{k}).
\end{equation}

In this paper, we assume that $U(\vec{r})$ is a short-range potential with
effective range $r_{\ast }$. In the region $r\equiv |\vec{r}|\gtrsim r_{\ast
}$, we have $U\left( \vec{r}\right) \simeq 0$ and further the low-energy
scattering state $|\Psi _{t}^{(+)}(\vec{r})\rangle $ with respect to the
incident state $|\Psi _{t}^{(0)}(\vec{r})\rangle $ can be expressed as~\cite%
{fewbody4,petrov01}
\begin{equation}
|\Psi _{t}^{(+)}(\vec{r})\rangle \approx |\Psi _{t}^{(0)}(\vec{r})\rangle
\!+B_{t}G_{0}\left( E_{t};\vec{r},0\right) |{\mathrm{S}}\rangle \,,
\label{sw2}
\end{equation}%
with $|{\mathrm{S}}\rangle =\left( |\!\!\uparrow \rangle _{1}|\!\!\downarrow
\rangle _{2}-|\!\!\downarrow \rangle _{1}|\!\!\uparrow \rangle _{2}\right) /%
\sqrt{2}$ the singlet spin state. Here $B_{t}$ is a $\vec{r}$-independent
constant and related with $|\Psi _{t}^{(+)}(\vec{r})\rangle $ through the
relation
\begin{equation}
\int d\vec{r}^{\,\prime }U(\vec{r}^{\,\prime })|\Psi _{t}^{(+)}(\vec{r}%
^{\,\prime })\rangle =B_{t}|\mathrm{S}\rangle .  \label{btr}
\end{equation}%
For details, see Appendix A. In Eq.~(\ref{sw2}) the free Green's function $%
G_{0}\left( \eta ;\vec{r},\vec{r}^{\,\prime }\right) $ is defined as
\begin{equation}
G_{0}\left( \eta ;\vec{r},\vec{r}^{\,\prime }\right) =\frac{1}{\eta
+i0^{+}\!-\!H_{0}\!}\delta \left( \vec{r}-\vec{r}^{\,\prime }\right) ,
\label{g2d}
\end{equation}%
and is a $\left( \vec{r},\vec{r}^{\,\prime }\right) $-dependent operator for
the two-atom spin. In this paper we consider the low-energy case with $k\ll
1/r_{\ast }$, and further assume that the SO coupling is weak enough so that
$\lambda <<1/r_{\ast }$. Furthermore, we have proved~\cite{ourbp} that, in
the \textit{short-range region} $r_{\ast }\lesssim r\ll 1/k$ the function $%
|\Psi _{t}^{(+)}(\vec{r})\rangle $ behaves as
\begin{equation}
|\Psi _{t}^{(+)}(\vec{r})\rangle \propto \left( \frac{1}{r}-\frac{1}{a_{%
\mathrm{\scriptscriptstyle R}}}\right) |\mathrm{S}\rangle -i\frac{\lambda }{2%
}\vec{c}\cdot \left( \frac{\vec{r}}{r}\right) |\mathrm{S}\rangle .
\label{src2d}
\end{equation}%
Here, the scattering length $a_{\mathrm{\scriptscriptstyle R}}$ is
determined by both the detail of the potential $U(\vec{r})$ and the SO
coupling. In some special systems, e.g., the systems with $[U(\vec{r}),\vec{c%
}]=0$ and those in the current experiments~\cite%
{NIST,NIST_elec,SOC,collective_SOC,JingPRA,NIST_partial,ourdecay}, the
scattering length $a_{\mathrm{\scriptscriptstyle R}}$ is independent of the
SO coupling and takes the same value as the scattering length in the systems
without SO coupling~\cite{ourbp}.

We can now calculate the the coefficient $B_{t}$ in Eq.~(\ref{sw2}) and the
inter-atomic scattering amplitude. As in Refs.~\cite{fewbody4,petrov01}, $%
B_{t}$ can be obtained from Eq.~(\ref{src2d}) and the behavior of $%
G_{0}\left( \eta ;\vec{r},\vec{r}^{\,\prime }\right) |{\mathrm{S}}\rangle $
in the short-range region $r_{\ast }\lesssim r<<1/k$. With calculations
shown in Appendix B, we find that%
\begin{eqnarray}
G_{0}\left( \eta ;\vec{r},0\right)  &\approx &-\frac{1}{4\pi }\left( \frac{1%
}{r}+i\eta ^{1/2}\right) +F\left( \eta \right) +i\frac{\lambda }{8\pi }\vec{c%
}\cdot \left( \frac{\vec{r}}{r}\right)   \notag \\
&&\ \ \ \ \ \ \ \ \ \ \ \ \ \ \ \ \ \ \ (\mbox{for }r_{\ast }\lesssim
r<<1/k)\,,  \label{gga}
\end{eqnarray}%
where the operator $F\left( \eta \right) $ is defined as
\begin{equation}
F\left( \eta \right) =\frac{1}{\left( 2\pi \right) ^{3}}\int d\vec{k}%
^{\prime \prime }\mathcal{F}(\eta ,\vec{k}^{\prime \prime })\,,
\label{ffeta}
\end{equation}%
where
\begin{equation}
\mathcal{F}(\eta ,\vec{k}^{\prime \prime })=\sum_{\alpha ^{\prime \prime
}}\left( \frac{|\alpha ^{\prime \prime },\vec{k}^{\prime \prime }\rangle
\langle \alpha ^{\prime \prime },\vec{k}^{\prime \prime }|}{\eta
+i0^{+}-E_{t^{\prime \prime }}}-\frac{|\alpha ^{\prime \prime },\vec{k}%
^{\prime \prime }\rangle \langle \alpha ^{\prime \prime },\vec{k}^{\prime
\prime }|}{\eta +i0^{+}-k^{\prime \prime 2}}\right) ,  \label{if}
\end{equation}%
with $t^{\prime \prime }=(\alpha ^{\prime \prime },\vec{K},\vec{k}^{\prime
\prime })$\ and $k^{\prime \prime }=|\vec{k}^{\prime \prime }|$. It is clear
that the fact $\lim_{|\vec{k}|\rightarrow \infty }h_{0}(\vec{k})=\lambda
\vec{c}\cdot \vec{k}$\ gives $\lim_{k^{\prime \prime }\rightarrow \infty
}[\mathcal{F}(\eta ,\vec{k}^{\prime \prime })+\mathcal{F}(\eta ,-\vec{k}^{\prime \prime
})]\propto 1/k^{\prime \prime 4}+O(1/k^{\prime \prime 5})$. Therefore, using
$\int d\vec{k}^{\prime \prime }=\int_{0}^{\infty }k^{\prime \prime
2}dk^{\prime \prime }\int d\Omega _{\vec{k}^{\prime \prime }}$\ with $\Omega
_{\vec{k}^{\prime \prime }}$\ the solid angle of $\vec{k}^{\prime \prime }$,
it is easy to prove that the integration in the r.h.s of Eq.~(\ref{ffeta})
converges to a finite operator. Apparently, $\mathcal{F}\left( \eta \right) $\ can be
re-expressed as $\mathcal{F}\left( \eta \right) =\left( 2\pi \right) ^{-3}\int d\vec{k}%
^{\prime \prime }[\mathcal{F}(\eta ,\vec{k}^{\prime \prime })+\mathcal{F}(\eta ,-\vec{k}^{\prime
\prime })]/2$. Such an expression maybe convenient for the numerical
calculation of the integration.

Substituting Eq.~(\ref{gga}) into Eq.~(\ref{sw2}), we get the expression for
$|\Psi _{t}^{(+)}(\vec{r})\rangle $ in the short-range region%
\begin{eqnarray}
|\Psi _{t}^{(+)}(\vec{r})\rangle =|\Psi _{t}^{(0)}(0)\rangle \!+iB_{t}\frac{%
\lambda }{8\pi }\vec{c}\cdot \left( \frac{\vec{r}}{r}\right) |{\mathrm{S}}%
\rangle &&  \notag \\
+B_{t}\left[ -\frac{1}{4\pi }\left( \frac{1}{r}+iE_{t}^{1/2}\right) +F\left(
E_{t}\right) \right] |{\mathrm{S}}\rangle &&  \notag \\
\ \ \ \ \ \ \ \ \ \ \ \ \ \ \ \ \ \ \ \ \ \ \ \ \ \ \ \ \ \ \ (\mbox{for }%
r_{\ast }\lesssim r<<1/k) &&.  \label{sea}
\end{eqnarray}%
Comparing Eq.~(\ref{sea}) to Eq.~(\ref{src2d}) and using the fact that $%
F(\eta )|{\mathrm{S}}\rangle \propto |{\mathrm{S}}\rangle $, we obtain
\begin{equation}
B_{t}=\frac{4\pi \!\langle {\mathrm{S}}|\Psi _{t}^{(0)}(0)\rangle }{1/a_{%
\mathrm{\scriptscriptstyle R}}+iE_{t}^{1/2}-4\pi \langle {\mathrm{S}}%
|F(E_{t})|{\mathrm{S}}\rangle }.  \label{at}
\end{equation}

According to scattering theory~\cite{scbook}, the scattering amplitude $%
f(t^{\prime }\leftarrow t)$ between the incident state $|\Psi _{t}^{(0)}(%
\vec{r})\rangle $ and an energy-conserved output state $|\Psi _{t^{\prime
}}^{(0)}(\vec{r})\rangle $ with $t^{\prime }=(\alpha ^{\prime },\vec{K},\vec{%
k}^{\prime })$ is defined as
\begin{equation}
f(t^{\prime }\leftarrow t)=-2\pi ^{2}\int d\vec{r}\langle \Psi _{t^{\prime
}}^{(0)}(\vec{r})|U(\vec{r})|\Psi _{t}^{(+)}\left( \vec{r}\right) \rangle .
\end{equation}%
It is pointed out that, since $U(\vec{r})\simeq 0$ in the region $r\gtrsim
r_{\ast }$, the integration in the right-hand side of the above equation is
only done in the region $r\lesssim r_{\ast }$. Under the low-energy
condition $k<<1/r_{\ast }$, in this region we have $\langle \Psi _{t^{\prime
}}^{(0)}(\vec{r})|\approx \langle \Psi _{t^{\prime }}^{(0)}(0)|$, and then $%
f(t^{\prime }\leftarrow t)$ can be re-expressed as
\begin{eqnarray}
f(t^{\prime }\leftarrow t) &=&-2\pi ^{2}\langle \Psi _{t^{\,\prime
}}^{(0)}(0)|\int d\vec{r}\,U(\vec{r})|\Psi _{t}^{(+)}\left( \vec{r}\right)
\rangle \,  \notag \\
&=&-2\pi ^{2}\!\langle \Psi _{t^{\prime }}^{(0)}(0)|\mathrm{S}\rangle
B_{t}\,,  \label{c2}
\end{eqnarray}%
where we have used Eq.~(\ref{btr}). Thus, using Eq.~(\ref{at}), we finally
have
\begin{eqnarray}
&&f(t^{\prime }\leftarrow t)  \notag \\
&=&-\left( 2\pi \right) ^{3}\!\langle \Psi _{t^{\prime }}^{(0)}(0)|\frac{1}{%
1/a_{\mathrm{\scriptscriptstyle R}}+iE_{t}^{1/2}-4\pi F(E_{t})}|\Psi
_{t}^{(0)}\left( 0\right) \rangle \,,  \notag \\
&&  \label{ff}
\end{eqnarray}%
with $E_{t}^{1/2}\equiv i\sqrt{|E_{t}|}$ for $E_{t}<0$. This is the exact
analytical expression for the low-energy scattering amplitude of two spin-$%
1/2$ fermonic atoms with SO coupling.

\subsection{Atoms with arbitrary spin}

The above approach can be straightforwardly generalized to the general case
of two fermonic or bosonic atoms with any kind of SO coupling and arbitrary
spin. In these cases, the single-atom motion and the relative motion of the
two atoms are still given by Eqs. (\ref{h1b}) and (\ref{hf}), respectively.
The incident state can be expressed as
\begin{equation}
|\Psi _{t}^{\left( 0\right) }(\vec{r})\rangle \!=\!\frac{e^{i\vec{k}\cdot
\vec{r}}}{4\pi ^{3/2}}|\alpha ,\vec{k}\rangle \pm \!\frac{e^{-i\vec{k}\cdot
\vec{r}}}{4\pi ^{3/2}}\mathrm{P}_{12}|\alpha ,\vec{k}\rangle \,,
\end{equation}%
where $\pm $ are for the systems of bosonic an fermonic atoms, respectively.

The scattering wave function with respect to incident state $|\Psi
_{t}^{\left( 0\right) }(\vec{r})\rangle $ can still be denoted as $|\Psi
_{t}^{\left( +\right) }(\vec{r})\rangle $. In the region with $r\gtrsim
r_{\ast }$ we have (Appendix A)%
\begin{equation}
|\Psi _{t}^{(+)}(\vec{r})\rangle \approx |\Psi _{t}^{(0)}(\vec{r})\rangle
\!+G_{0}\left( E_{t};\vec{r},0\right) |\phi \rangle \,  \label{21}
\end{equation}%
with the $\vec{r}$-independent spin state $|\phi \rangle $ satisfies

\begin{equation}
|\phi \rangle =\int d\vec{r}^{\,\prime }U(\vec{r}^{\,\prime })|\Psi
_{t}^{(+)}(\vec{r})\rangle .  \label{btr2}
\end{equation}%
This is very similar to Eq.~(\ref{sw2}), but the spin state $|\phi \rangle $
is not unique. Instead, $|\phi \rangle $ can be different for different
incident state $|\Psi _{t}^{\left( 0\right) }(\vec{r})\rangle $.
Furthermore, as shown in Ref.~\cite{ourbp}, in the short-range region $%
r_{\ast }\lesssim r<<1/k$ the scattering wave function $|\Psi _{t}^{\left(
+\right) }(\vec{r})\rangle \!$ behaves as%
\begin{equation}
|\Psi _{t}^{\left( +\right) }(\vec{r})\rangle \!=\left( \frac{1}{r}-A_{%
\mathrm{R}}\right) |\chi \rangle -i\frac{\lambda }{2}\vec{c}\cdot \left(
\frac{\vec{r}}{r}\right) |\chi \rangle .  \label{22}
\end{equation}%
Here $|\chi \rangle $ is another $\vec{r}$-independent spin state, and $A_{%
\mathrm{R}}$ is a $\vec{r}$-independent operator in the spin space, which is
also determined by the detail of the interaction potential $U(\vec{r})$ and
the SO coupling. For the cases of spin$-1/2$ fermonic atoms, we have $A_{%
\mathrm{R}}=1/a_{\mathrm{R}}$, and Eq.~(\ref{22}) reduces to Eq.~(\ref{src2d}%
). As in the above section, in the current experiments~\cite%
{NIST,NIST_elec,SOC,collective_SOC,JingPRA,NIST_partial,ourdecay} for
bosonic atoms with 1D SO coupling, $A_{\mathrm{R}}$ is independent of the SO
coupling. For instance, for the ultracold gases with spin-$1$ $^{87}$Rb
atoms, we have%
\begin{equation}
A_{\mathrm{R}}=\frac{1}{a_{0}}\mathcal{P}_{F=0}+\frac{1}{a_{2}}\mathcal{P}%
_{F=2},
\end{equation}%
where $a_{0}$ $(a_{2})$ is the scattering length with respect to the total
atomic spin $F=0$ ($F=2$) and $\mathcal{P}_{F=0,2}$ are the associated
projection operators.

Using Eq.~(\ref{gga}), we can obtain the expression of $|\Psi _{t}^{\left(
+\right) }(\vec{r})\rangle \!$ in the short-range region. Comparing such an
expression with Eq.~(\ref{22}), we have%
\begin{equation}
|\phi \rangle =-4\pi |\chi \rangle =\frac{4\pi }{A_{\mathrm{R}%
}+iE_{t}^{1/2}-4\pi F(E_{t})}|\Psi _{t}^{(0)}(0)\rangle  \label{ka}
\end{equation}%
and the exact analytical expression%
\begin{eqnarray}
&&f(t^{\prime }\leftarrow t)  \notag \\
&=&-\left( 2\pi \right) ^{3}\!\langle \Psi _{t^{\prime }}^{(0)}(0)|\frac{1}{%
A_{\mathrm{R}}+iE_{t}^{1/2}-4\pi F(E_{t})}|\Psi _{t}^{(0)}\left( 0\right)
\rangle \,.  \notag \\
&&  \label{fa}
\end{eqnarray}%
for the scattering amplitude of SO-coupled ultracold atoms with arbitrary
spin. In Ref. \cite{ourdecay} we have used this result to quantitatively explain
the collisional decay observed in our experiments with SO-coupled $^{87}$Rb atoms.

It is pointed out that, in the presence of SO-coupling, the state $|\Psi
_{t}^{(0)}\left( 0\right) \rangle $ depends on the quantum number $t=(\alpha
,\vec{K},\vec{k})$, and thus changes with the direction of the atomic
relative momentum $\vec{k}$. Therefore, the scattering amplitude $%
f(t^{\prime }\leftarrow t)$ in Eq.~(\ref{fa}) is anisotropic with respect to
the directions of the incident momentum $\vec{k}$ and output momentum $\vec{%
k^{\prime }}$. This anisotropicity is also observed in the experiment by I.
B. Spielman \textit{et. al.}~\cite{NIST_partial}.

In the end of this section we consider a simple case that $A_{\mathrm{R}%
}=1/a $ with $a$ a constant c-number and $|a|$ is much smaller than the
eigen-values of the operator $iE_{t}^{1/2}-4\pi F(E_{t})$. In this case Eq.~(%
\ref{fa}) can be simplified as $f(t^{\prime }\leftarrow t)\approx -a\left(
2\pi \right) ^{3}\!\langle \Psi _{t^{\prime }}^{(0)}(0)|\Psi
_{t}^{(0)}\left( 0\right) \rangle $. This approximate result can also be
obtained with Fermi's golden rule. On the other hand, if the eigen-values of
$A_{\mathrm{R}}$ is comparable to or smaller than the ones of the operator $%
iE_{t}^{1/2}-4\pi F(E_{t})$, the contribution from that operator becomes
significant and the Fermi's golden rule is no longer applicable.

\subsection{Two-atom bound state}

In the above subsections we derived the analytical expression of the
scattering amplitude of two ultra-cold atoms with SO-coupling. The Green's
function approach used in our calculations can also be applied to derive the
low-energy bound state of two SO-coupled atoms. We denote the energy of the
bound state as $E_{b}$, then the binding energy can be defined as $E_{%
\mathrm{binding}}\equiv -(E_{b}-E_{\mathrm{th}})$, where $E_{\mathrm{th}}$
is the energy of scattering threshold or the lowest eigen-energy of $H_{0}$.
As shown in Appendix A, when $E_{\mathrm{binding}}<<r_{\ast }^{-2}$, the
wave function $|\Psi _{b}(\vec{r})\rangle $ of low-energy bound state is
given by
\begin{equation}
|\Psi _{b}(\vec{r})\rangle \approx N_{b}G_{0}\left( E_{b};\vec{r},0\right)
|\phi _{b}\rangle \,,  \label{psib}
\end{equation}%
in the region $r\gtrsim r_{\ast }$. Here $N_{b}$ is the normalization factor
and $|\phi _{b}\rangle \,$\ is a $\vec{r}$-independent spin state.
Furthermore, in the short-range region $r_{\ast }\lesssim r<<E_{\mathrm{%
binding}}^{-1/2}$, $|\Psi _{b}(\vec{r})\rangle $ has the same behavior as
the low-energy scattering state $|\Psi _{t}^{(+)}(\vec{r})\rangle $ and thus
can be expressed as $|\Psi _{b}(\vec{r})\rangle =(1/r-A_{\mathrm{R}})|\chi
_{b}\rangle -i(\lambda /2)\vec{c}\cdot (\vec{r}/r)|\chi _{b}\rangle $, with $%
|\chi _{b}\rangle $ a $\vec{r}$-indepedent state in the spin space. As in
the above section, with this fact and the short-range behavior of $%
G_{0}\left( \eta ;\vec{r},0\right) $ given by Eq.~(\ref{gga}), we find that $%
|\chi _{b}\rangle $, $|\phi _{b}\rangle $ and $E_{b}$ can be obtained by
\begin{eqnarray}
\left[ -iE_{b}^{1/2}+4\pi F\left( E_{b}\right) \right] |\chi _{b}\rangle
&=&A_{\mathrm{R}}|\chi _{b}\rangle ,\   \label{eeb} \\
|\chi _{b}\rangle &=&-\frac{1}{4\pi }N_{b}|\phi _{b}\rangle ,
\end{eqnarray}%
with $E_{b}^{1/2}\equiv i\sqrt{|E_{b}|}$for $E_{b}<0$. In particular, for
two spin$-1/2$ fermonic atoms, we have $A_{\mathrm{R}}=1/a_{\mathrm{%
\scriptscriptstyle R}}$ and $|\phi _{b}\rangle =|\mathrm{S}\rangle $. Then
Eq.~(\ref{eeb}) becomes
\begin{equation}
-iE_{b}^{1/2}+4\pi \langle \mathrm{S}|F\left( E_{b}\right) |\mathrm{S}%
\rangle =\frac{1}{a_{\mathrm{\scriptscriptstyle R}}}.  \label{eeb2}
\end{equation}%
With this equation one can obtain the bound-state energy $E_{b}$.
\begin{figure}[tbp]
\centering
\subfigure{
    \includegraphics[bb=61 234 485 571,clip,width=7.5cm]{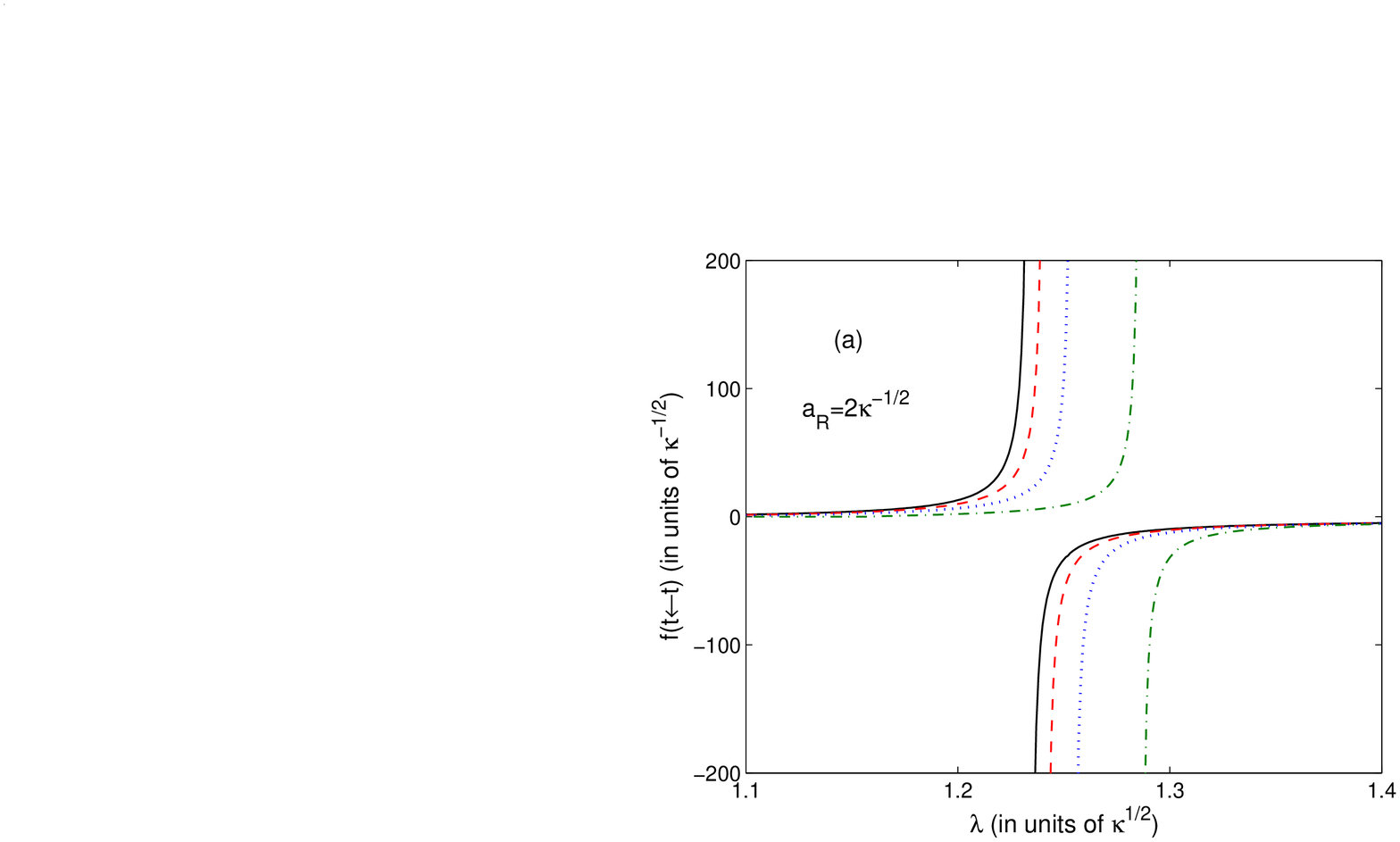}}
\subfigure{
    \includegraphics[bb=61 234 485 571,clip, width=7.5cm]{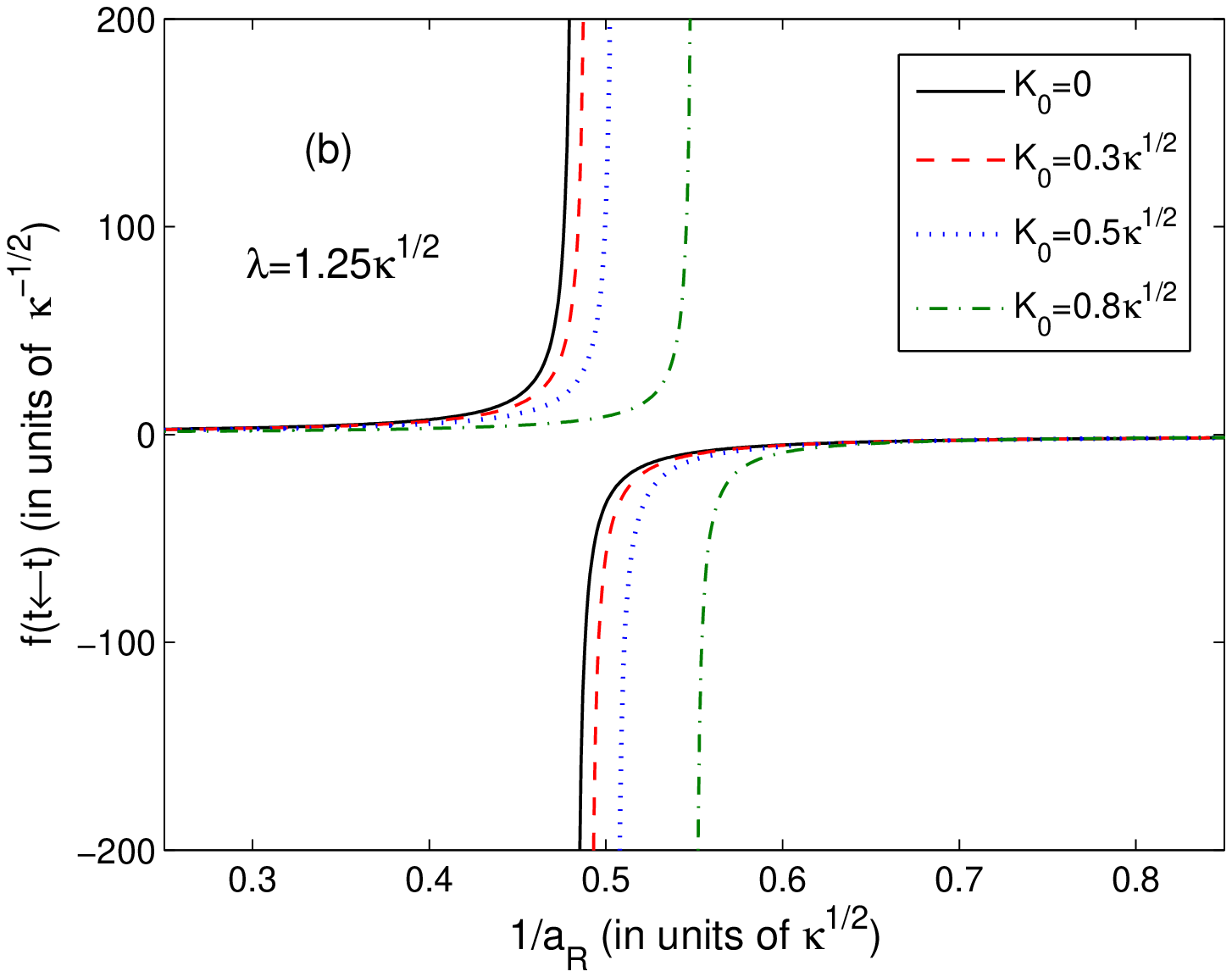}}
\caption{(Color online) The threshold scattering scattering amplitude $%
f(t\leftarrow t)$ of two spin-$1/2$ fermoinc atoms with 1D SO coupling. We
plot the $f(t\leftarrow t)$ as a function of SO-coupling intensity $\protect%
\lambda $ with the scattering length $a_{\mathrm{\scriptscriptstyle R}}=2%
\protect\kappa ^{-1/2}$ (a) and a function $a_{\mathrm{\scriptscriptstyle R}%
} $ with $\protect\lambda =1.25\protect\kappa ^{1/2}$ (b). Here we have used
the natural unit with $\hbar =m=1$ and set the total momentum of the two
atoms to be $\vec{K}=(K_{0},0,0)$, with $K_{0}=0$ (black solid line), $0.3%
\protect\kappa ^{1/2}$ (red dashed line), $0.5\protect\kappa ^{1/2}$ (green
dotted line) and $0.8\protect\kappa ^{1/2}$ (blue dashed-dotted line). }
\label{reso}
\end{figure}

\section{SO-coupling-induced resonance}

In above discussions, we derive the general analytical expression (\ref{fa})
for the scattering amplitude of ultracold atoms with SO coupling. As an
application, in this section we will show that SO coupling can induce
scattering resonance of two ultracold atoms. For simplicity, here we only
consider the scattering of two spin-$1/2$ fermoinc atoms.

In such a system, the incident state $|\Psi _{t}^{(0)}(\vec{r})\rangle $
takes the form in Eq.~(\ref{psi0}), and one has $|\Psi _{t}^{(0)}(0)\rangle
\propto |\mathrm{S}\rangle $. Thus, the inter-atomic scattering amplitude $%
f(t^{\prime }\leftarrow t)$ in Eq.~(\ref{ff}) can be re-written as%
\begin{equation}
f(t^{\prime }\leftarrow t)=-\left( 2\pi \right) ^{3}\!\frac{\langle \Psi
_{t^{\prime }}^{(0)}(0)|\mathrm{S}\rangle \langle \mathrm{S}|\Psi
_{t}^{(0)}\left( 0\right) \rangle }{1/a_{\mathrm{\scriptscriptstyle R}%
}+d(\lambda ,E_{t})},  \label{ff2}
\end{equation}%
with the function $d(\lambda ,E_{t})$ defined as
\begin{equation}
d(\lambda ,E_{t})=iE_{t}^{1/2}-4\pi \langle \mathrm{S}|F(E_{t})|\mathrm{S}%
\rangle .
\end{equation}%
We first consider the case of threshold scattering with $E_{t}=E_{\mathrm{th}%
}$. In this case, we have $|\Psi _{t^{\prime }}^{(0)}\left( \vec{r}\right)
\rangle =|\Psi _{t}^{(0)}\left( \vec{r}\right) \rangle $ and $d(\lambda ,E_{%
\mathrm{th}})$ usually takes a real value. Therefore, when the condition
\begin{equation}
\frac{1}{a_{\mathrm{\scriptscriptstyle R}}}+d(\lambda ,E_{\mathrm{th}})=0
\label{rc}
\end{equation}%
is satisfied, the threshold scattering amplitude diverges and a scattering
resonance occurs. Since the scattering length $a_{\mathrm{\scriptscriptstyle %
R}}$ can be finite, such a resonance is induced by the SO coupling.
Comparing Eq.~(\ref{rc}) to Eq.~(\ref{eeb2}), we further find that a bound
state with zero binding energy appears at the resonance point. In
experiments, one can observe the SO-coupling-induced resonance by tuning the
scattering length or the SO-coupling intensity~\cite{chuanweizhang}.

In Fig.~\ref{reso} we plot the amplitude $f(t\leftarrow t)$ of the threshold
scattering of two spin-$1/2$ fermoinc atoms with 1D SO coupling, i.e., $\vec{%
M}=(\hat{\sigma}_{z},0,0)$ and $Z=\kappa \hat{\sigma}_{x}$, as in current
experiments. In our calculation we assume the total momentum $\vec{K}$ of
the two atoms is along the $x$-direction, i.e., $\vec{K}=(K_{0},0,0)$. The
scattering amplitudes with respect to different values of $K_{0}$ are
illustrated versus $1/a_{\mathrm{\scriptscriptstyle R}}$ and the SO coupling
intensity $\lambda $. The appearance of resonance is clearly shown.
\begin{figure}[tbp]
\includegraphics[bb=92 233 470 572,clip,width=7.5cm]{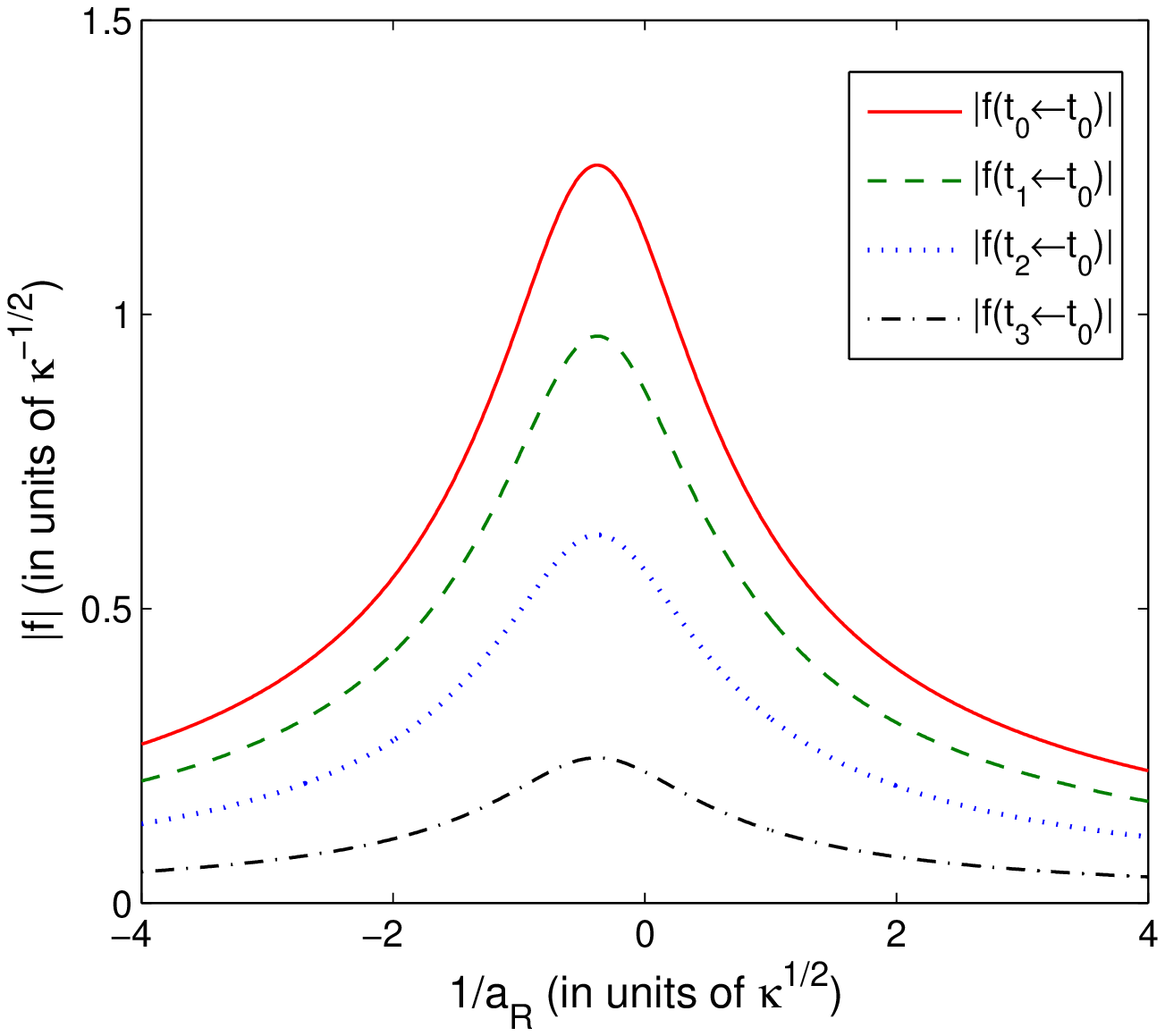}
\caption{(Color online) The absolute value of scattering amplitude of two
spin-$1/2$ fermonic atoms with 1D SO-coupling. Here we plot $|f(t^{\prime
}\longleftarrow t)|$ for the cases with $\vec{K}=0$, $\protect\lambda =1.25%
\protect\kappa ^{1/2}$ and $E_{t}>E_{\mathrm{th}}$. We take $t=(2,0,0)$ and $%
t^{\prime }=t_{0}$ (blue solid line), $t_{1}$ (green dashed line), $t_{2}$
(red dotted line) and $t_{3}$ (black dashed-dotted line), with $t_{0,1,2,3}$
defined in Sec. III.}
\label{fbigk}
\end{figure}

Next, when the scattering energy $E_{t}$ is larger than the scattering
threshold, the function $d(\lambda ,E_{t})$ takes a complex value. Thus, the
scattering amplitude $f(t^{\prime }\leftarrow t)$ cannot divergent.
Nevertheless, as a function of $1/a_{\mathrm{\scriptscriptstyle R}}$, the
absolute value of the scattering amplitude still achieves a local maximum
value when the condition $\mathrm{Re}[1/a_{\mathrm{\scriptscriptstyle R}%
}+d(\lambda ,E_{t})]=0$ or $1/a_{\mathrm{\scriptscriptstyle R}}=-\mathrm{Re}%
[d(\lambda ,E_{t})]$ is satisfied. To illustrate this effect, we also
calculate the scattering amplitude $f(t^{\prime }\leftarrow t)$ for two spin-%
$1/2$ fermoinc atoms with $\vec{M}=(\hat{\sigma}_{z},0,0)$ and $Z=\kappa
\hat{\sigma}_{x}$. In Fig.~\ref{fbigk} we plot $|f(t^{\prime }\leftarrow t)|$
versus $1/a_{\mathrm{\scriptscriptstyle R}}$ for the cases with $\vec{K}=0$,
$t=(2,0,0)$ and $t^{\prime }=t_{0,1,2,3}$ where $t_{0}=t$, $t_{1}=(4,0,\sqrt{%
\lambda ^{4}-\kappa ^{2}}/\lambda )$, $t_{2}=(4,0,\sqrt{\lambda ^{4}-\kappa
^{2}}/\lambda -0.5\kappa ^{1/2})$ and $t_{3}=(4,0,\sqrt{\lambda ^{4}-\kappa
^{2}}/\lambda -0.8\kappa ^{1/2})$. The peak behavior of the scattering
amplitude is clearly illustrated.

We mention again that the resonance discussed here is essentially induced by
the SO-coupling term in the Hamiltonian $H_{0}$ defined in Eq.~(\ref{hf}).
From Eq.~(\ref{hf}) one can easily find that such a term can be omitted when
$k>>\lambda $, and therefore the SO-coupling-induced resonance is
significant only when the atomic relative momentum $k$ is small enough i.e.,
$k\lesssim \lambda $. Finally, we emphasize that the calculations in this
seciton can be directly generalized to the systems of atoms with arbitrary
spin and SO coupling, and the SO-coupling-induced resonance can also appear.

\section{Effective interactions}

In the many-body theory of ultracold gases, the inter-atomic interaction is
usually modeled by some simple effective potentials. The most widely-used
effective interactions include the Huang-Yang pseudo potential~\cite{hyp},
the Bethe-Peierls boundary condition~\cite{bpc} and the contact interaction
with a momentum cutoff~\cite{pr}.

For a given system, the inter-atomic scattering amplitude given by the
effective inter-atomic interaction should be the same as the one from the
realistic interaction potential $U(\vec{r})$. In our previous works~\cite%
{fewbody4,ourbp}, we have shown that to satisfy such a condition, the
Bethe-Peierls boundary condition for 3D ultracold gases, as well as the
renormalization relation for the contact interaction of quasi-2D gases,
should be modified in the presence of SO coupling. In this section, we
consider the contact interaction with a momentum cutoff in a 3D and pure-2D
uniform system of the SO-coupled spin-$1/2$ Fermi gas, as well as the
Huang-Yang pseudo potential.

\subsection{Contact interaction in 3D system}

In this and the next subsection we give up the $\vec{r}$-representation, and
use the Dirac symbol $|\rangle \rangle $\ defined in Sec. II.A to describe
the state in the total Hilbert space $H$. In a 3D system, the contact
interaction $\hat{U}_{\mathrm{eff}}$\ with a momentum cutoff can be
expressed as an operator in $H$:%
\begin{equation}
\hat{U}_{\mathrm{eff}}=\frac{U_{0}}{\left( 2\pi \right) ^{3}}%
\int_{\left\vert \vec{k}\right\vert ,\left\vert \vec{k}^{\prime }\right\vert
<k_{c}}|\vec{k})(\vec{k}^{\prime }|\otimes |\mathrm{S}\rangle \langle
\mathrm{S}|d\vec{k}d\vec{k}^{\prime }.
\end{equation}%
Here $k_{c}$\ is a cut-off momentum and $|\vec{k})$\ $\equiv \left( 2\pi
\right) ^{-3/2}\int d\vec{r}e^{i\vec{k}\cdot \vec{r}}|\vec{r})$\ is a state
in the space $H_{r}$.\ It is pointed out that, in many references about the
many-body theory of ultracold gases, the systems are first assumed to have
finite volume ${\cal V}$, and the final result is obtained in the limit $%
{\cal V}\rightarrow \infty $. In these cases the second-quantized form $\hat{U}_{%
\mathrm{eff}}$\ is given by $\hat{U}_{\mathrm{eff}}=U_{0}/{\cal V}\sum_{%
\vec{k},\vec{k}^{\prime },\vec{K}}\,^{\prime }a_{\vec{K}/2+\vec{k},\uparrow
}^{\dagger }a_{\vec{K}/2-\vec{k},\downarrow }^{\dagger }a_{\vec{K}/2-\vec{k}%
^{\prime },\downarrow }a_{\vec{K}/2+\vec{k}^{\prime },\uparrow }$, where $a_{%
\vec{p},\sigma }^{\dagger }$\ and $a_{\vec{p},\sigma }$\ are the creation
and annihilation operators for an atom with momentum $\vec{p}$\ and spin $%
\sigma $. The summation $\sum_{\vec{k},\vec{k}^{\prime },\vec{K}}^{\prime }$%
\ is under the condition $\max (|\vec{k}|,|\vec{k}^{\prime }|)<k_{c}$\ with $%
k_{c}$\ is a cut-off momentum.

The renormalization relation for this contact potential, i.e., the
relationship between $U_{0}$ and $k_{c}$, can be obtained from the condition
that $\hat{U}_{\mathrm{eff}}$ and the realistic inter-atomic interaction
should lead to the same low-energy scattering amplitude. For the cases
without SO-coupling, the standard calculation gives the well-known result%
\begin{equation}
\frac{1}{4\pi a_{s}}=\frac{1}{U_{0}}+\frac{1}{(2\pi )^{3}}\int_{k^{\prime
\prime }<k_{c}}d\vec{k}^{\prime \prime }\frac{1}{k^{\prime \prime 2}},
\label{rn}
\end{equation}%
with $a_{s}$ the $s$-wave scattering length.

In the presence of SO coupling, the correct renormalization relation can be
obtained by the same procedure. The effective scattering amplitude $f_{%
\mathrm{eff}}(t^{\prime }\leftarrow t)$ given by $\hat{U}_{\mathrm{eff}}$
can be obtained from the Lippmman-Schwinger equation for the two-body $T$%
-operator $\hat{T}_{\mathrm{eff}}\left( \eta \right) $ with respect to $\hat{%
U}_{\mathrm{eff}}$:%
\begin{equation}
\hat{T}_{\mathrm{eff}}\left( \eta \right) =\hat{U}_{\mathrm{eff}}+\hat{U}_{%
\mathrm{eff}}\hat{G}_{0}\left( \eta \right) \hat{T}_{\mathrm{eff}}\left(
\eta \right) ,  \label{ls}
\end{equation}%
where the Green's operator $\hat{G}_{0}\left( \eta \right) $ is defined as%
\begin{eqnarray}
\hat{G}_{0}\left( \eta \right) &=&\frac{1}{\eta +i0^{+}\!-\!H_{0}\!} \\
&=&\sum_{\alpha }\int d\vec{k}\frac{|\vec{k})(\vec{k}|\otimes |\alpha ,\vec{k%
}\rangle \langle \alpha ,\vec{k}|}{\eta +i0^{+}-E_{t}},
\end{eqnarray}%
where we have $t=(\alpha ,\vec{K},\vec{k})$ as before.

Using the Lippmann-Schwinger equation, we can obtain the equation for the $T$%
-matrix element:%
\begin{widetext}
\begin{eqnarray}
&&T_{\mathrm{eff}}\left( \eta ,t^{\prime },t\right) =\frac{U_{0}}{\left(
2\pi \right) ^{3}}\langle \alpha ^{\prime },\vec{k}^{\prime }|\mathrm{S}%
\rangle \langle \mathrm{S}|\alpha ,\vec{k}\rangle +\frac{U_{0}}{\left( 2\pi
\right) ^{3}}\langle \alpha ^{\prime },\vec{k}^{\prime }|\mathrm{S}\rangle %
\left( \sum_{\alpha ^{\prime \prime }}\int_{\left\vert \vec{k}^{\prime
\prime }\right\vert <k_{c}}d\vec{k}^{\prime \prime }\frac{\langle \mathrm{S}%
|\alpha ^{\prime \prime },\vec{k}^{\prime \prime }\rangle }{\eta
+i0^{+}-E_{t^{\prime \prime }}}T_{\mathrm{eff}}\left( \eta ,t^{\prime \prime
},t\right) \right) ,  \notag \\
&&  \label{lst}
\end{eqnarray}%
\end{widetext}where we have $t^{\prime }=(\alpha ^{\prime },\vec{K}^{\prime
},\vec{k}^{\prime })$ and $T_{\mathrm{eff}}\left( \eta ,t^{\prime },t\right)
=(\vec{k}^{\prime }|\langle \alpha ^{\prime },\vec{k}^{\prime }|\hat{T}_{%
\mathrm{eff}}\left( \eta \right) |\alpha ,\vec{k}\rangle |\vec{k})$. The
notations $t^{\prime \prime }$ and $T_{\mathrm{eff}}\left( \eta ,t^{\prime
\prime },t\right) $ are defined similarly. From Eq.~(\ref{lst}), we find
that $T_{\mathrm{eff}}\left( \eta ,t^{\prime },t\right) $ can be expressed as%
\begin{equation}
T_{\mathrm{eff}}\left( \eta ,t^{\prime },t\right) =\frac{U_{0}}{\left( 2\pi
\right) ^{3}}\langle \alpha ^{\prime },\vec{k}^{\prime }|\mathrm{S}\rangle
u\left( \eta ,\alpha ,\vec{k}\right)  \label{taa}
\end{equation}%
where $u(\eta ,\alpha ,\vec{k})$ is independent of $\alpha ^{\prime }$ and $%
\vec{k}^{\prime }$. Substituting Eq.~(\ref{taa}) into Eq.~(\ref{lst}), we
get the result
\begin{eqnarray}
u\left( \eta ,\alpha ,\vec{k}\right) &=&\frac{\langle \mathrm{S}|\alpha ,%
\vec{k}\rangle }{1-\frac{U_{0}}{\left( 2\pi \right) ^{3}}\left( \sum_{\alpha
^{\prime \prime }}\int_{k^{\prime \prime }<k_{c}}d\vec{k}^{\prime \prime }%
\frac{\left\vert \langle \mathrm{S}|\alpha ^{\prime \prime },\vec{k}^{\prime
\prime }\rangle \right\vert ^{2}}{\eta +i0^{+}-E_{t^{\prime \prime }}}%
\right) }.  \notag \\
&&  \label{tbb}
\end{eqnarray}%
Substituting Eq.~(\ref{tbb}) into Eq.~(\ref{taa}), we can obtain the
expression of the $T$-matrix element $T_{\mathrm{eff}}\left( \eta ,t^{\prime
},t\right) $.

The scattering amplitude $f_{\mathrm{eff}}(t^{\prime }\leftarrow t)$ for two
spin-$1/2$ fermionic atoms is defined as
\begin{equation}
f_{\mathrm{eff}}(t^{\prime }\leftarrow t)=-2\pi ^{2}\ \langle \langle \Psi
_{t^{\prime }}^{(0)}|\hat{T}_{\mathrm{eff}}\left( E_{t}\right) |\Psi
_{t}^{(0)}\rangle \rangle ,
\end{equation}%
where the incident state is $|\Psi _{t}^{(0)}\rangle \rangle
=2^{-1/2}[|\alpha ,\vec{k}\rangle |\vec{k})-(\mathrm{P}_{12}|\alpha ,\vec{k}%
\rangle )|-\vec{k})]$, with $\mathrm{P}_{12}$ the permutation operator for
the spin of the two atoms. Therefore, it is apparent that we have%
\begin{equation}
f_{\mathrm{eff}}(t^{\prime }\leftarrow t)=-2\pi ^{2}\left[ T_{\mathrm{eff}%
}\left( E_{t},t^{\prime },t\right) -T_{\mathrm{eff}}^{\prime }\left(
E_{t},t^{\prime },t\right) \right]  \label{ffe}
\end{equation}%
with $T_{\mathrm{eff}}^{\prime }\left( E_{t},t^{\prime },t\right) =(\vec{k}%
^{\prime }|\langle \alpha ^{\prime },\vec{k}^{\prime }|\hat{T}_{\mathrm{eff}%
}\left( \eta \right) [\mathrm{P}_{12}|\alpha ,\vec{k}\rangle ]|-\vec{k})$.
Substituting our result of $T_{\mathrm{eff}}\left( E_{t},t^{\prime
},t\right) $ into Eq.~(\ref{ffe}), we have
\begin{eqnarray}
f_{\mathrm{eff}}(t^{\prime } &\leftarrow &t)=-\frac{\frac{U_{0}}{2\pi }%
\langle \alpha ^{\prime },\vec{k}^{\prime }|\mathrm{S}\rangle \langle
\mathrm{S}|\alpha ,\vec{k}\rangle }{1-\frac{U_{0}}{\left( 2\pi \right) ^{3}}%
\left( \sum_{\alpha ^{\prime \prime }}\int_{k^{\prime \prime }<k_{c}}d\vec{k}%
^{\prime \prime }\frac{\left\vert \langle \mathrm{S}|\alpha ^{\prime \prime
},\vec{k}^{\prime \prime }\rangle \right\vert ^{2}}{\eta
+i0^{+}-E_{t^{\prime \prime }}}\right) }.  \notag \\
&&
\end{eqnarray}

By requiring that the zero-energy effective scattering amplitude $f_{\mathrm{%
eff}}(t^{\prime }\leftarrow t)$ be equal to the realistic scattering
amplitude, i.e.,
\begin{equation}
f_{\mathrm{eff}}(t^{\prime }\leftarrow t)=f(t^{\prime }\leftarrow t)
\end{equation}%
with $E_{t}=0$ and $f(t^{\prime }\leftarrow t)$ given by Eq.~(\ref{ff}), we
obtain the following renormalization relation for systems with SO coupling
in the limit $k_{c}\rightarrow \infty $:%
\begin{equation}
\frac{1}{4\pi a_{\mathrm{\scriptscriptstyle R}}}=\frac{1}{U_{0}}+\frac{1}{%
(2\pi )^{3}}\int_{k^{\prime \prime }<k_{c}}d\vec{k}^{\prime \prime }\frac{1}{%
|\vec{k}^{\prime \prime }|^{2}}.  \label{rna}
\end{equation}%
Here we have used Eq.~(\ref{bb12}). Comparing Eq.~(\ref{rn}) with Eq.~(\ref%
{rna}), we find that for the 3D contanct potential with momentum cutoff, the
form of the renormalization relation is not changed by the SO coupling. In
the presnece of SO coupling one only needs to replace the scattering length $%
a_{s}$ with $a_{\mathrm{\scriptscriptstyle R}}$.

We mention that Eq.~(\ref{eeb}) for the bound-state energy $E_{b}$ can also
be obtained from the contact potential $\hat{U}_{\mathrm{eff}}$ with
renormalization relation (\ref{rna}).

\subsection{Contact potential in pure-2D system}

Our above discussion can be directly generalized to pure-2D ultracold gases
of spin-$1/2$ fermonic atoms with SO coupling. We assume the atoms are
moving in the $x-y$ plane. Thus, the single-atom Hamiltonian is also given
by Eq.~(\ref{h1b}), with $\vec{P}=(P_{x},P_{y})$ the single-atom momentum in
the $x-y$ plane. The Hamiltonian for the relative motion is then%
\begin{equation}
H^{(2D)}=\vec{p}^{2}+\lambda \vec{c}{\cdot }\vec{p}+B(\vec{K})+U_{2D}(\vec{%
\rho})\equiv H_{0}^{(2D)}+U_{2D}(\vec{\rho}),
\end{equation}%
where $\vec{K}$ is the two-atom total momentum, $\vec{\rho}=(x,y)$ is the
two-atom relative position in the $x-y$ plane, and $U_{2D}$ is the two-atom
interaction potential in the 2D space, with effecitve range $\rho _{\ast }$.
Here the operators $\vec{c}$ and $B(\vec{K})$ in the two-atom spin space are
defined as in Sec. II, and SO-coupling intensity $\lambda $ is also assumed
to be small enough so that the condition $\lambda <<1/\rho _{\ast }$ is
satisfied.

Similar as in Sec. IIA, the incident state in the scattering, or the
eigen-state of the Hamiltonian $H_{0}^{(2D)}$, can be described by the
spinor wave function
\begin{equation}
|\Psi _{t}^{\left( 0\right) }(\vec{r})\rangle \!=\!\frac{e^{i\vec{k}\cdot
\vec{\rho}}}{2^{3/2}\pi }|\alpha ,\vec{k}\rangle -\!\frac{e^{-i\vec{k}\cdot
\vec{\rho}}}{2^{3/2}\pi }\mathrm{P}_{12}|\alpha ,\vec{k}\rangle \,
\end{equation}%
in the $\vec{r}$-representation. Here the state $|\alpha ,\vec{k}\rangle $
for the two-atom spin is defined as in Sec. IIA, and we have $t=(\alpha ,%
\vec{K},\vec{k})$ as before. In Ref. \cite{fewbody4}, we have calculated the
2D scattering amplitude $f^{\left( 2D\right) }(t^{\prime }\leftarrow t)$
between $|\Psi _{t}^{\left( 0\right) }(\vec{r})\rangle \!$ and $|\Psi
_{t^{\prime }}^{\left( 0\right) }(\vec{r})\rangle \!$ for the cases with
Rashba SO-coupling. The method applied there can be directly used for the
cases with arbitrary type of SO-coupling. The straightforward calculation
yields%
\begin{eqnarray}
&&f^{\left( 2D\right) }\left( t^{\prime }\longleftarrow t\right)  \notag \\
&=&-\frac{4\pi ^{3}\langle \Psi _{t^{\prime }}^{\left( 0\right) }(0)|\mathrm{%
S}\rangle \langle \mathrm{S}|\Psi _{t}^{\left( 0\right) }(0)\rangle }{\frac{%
i\pi }{2}-\ln E_{t}^{1/2}-C-\ln \frac{d_{R}}{2}-2\pi \mathcal{F}%
^{(2D)}\left( E_{t}\right) },  \label{f2d}
\end{eqnarray}%
where $C=0.5772...$ is the Euler $\mathrm{\Gamma }$ number, $E_{t}$ is the
eigen-value of $H_{0}^{(2D)}$, with respect to the eigen-state $|\Psi
_{t}^{\left( 0\right) }(\vec{r})\rangle $. In Eq.~(\ref{f2d}) the function $%
\mathcal{F}^{(2D)}\left( \eta \right) $ is defined as%
\begin{eqnarray}
&&\mathcal{F}^{(2D)}\left( \eta \right) =\frac{1}{\left( 2\pi \right) ^{2}}%
\sum_{\alpha ^{\prime \prime }}\int d\vec{k}^{\prime \prime }\left\vert
\langle \mathrm{S}|\alpha ^{\prime \prime },\vec{k}^{\prime \prime }\rangle
\right\vert ^{2}\times  \notag \\
&&\left( \frac{1}{\eta +i0^{+}-E_{t^{\prime \prime }}}-\frac{1}{\eta
+i0^{+}-|\vec{k}^{\prime \prime }|^{2}}\right) ,  \label{lam2d}
\end{eqnarray}%
with $t^{\prime \prime }=(\alpha ^{\prime \prime },\vec{K},\vec{k}^{\prime
\prime })$. The parameter $d_{R}$ in Eq.~(\ref{f2d}) can be determined by
the following condition (see Appendix B of Ref.~\cite{fewbody4} where $d_{R}$
is denoted by $d$): in the region with $\rho _{\ast }<|\vec{\rho}|<<1/k$,
the solution $|\psi _{R}(\vec{\rho})\rangle $ of equation%
\begin{equation}
\left[ \mathcal{T}(\vec{\rho})H^{(2D)}\mathcal{T}^{\dagger }(\vec{\rho})%
\right] |\psi _{T}(\vec{\rho})\rangle =E_{t}|\psi _{T}(\vec{\rho})\rangle
\end{equation}%
satisfies $|\psi _{T}(\vec{\rho})\rangle \propto (\ln |\vec{\rho}|-\ln
d_{R})|\mathrm{S}\rangle $. Here the rotation $\mathcal{T}(\vec{\rho})$ is
defined as $\mathcal{T}(\vec{\rho})=\exp (i\lambda c_{x}x/2)\exp (i\lambda
c_{y}y/2)$.

Now we consider the 2D contact potential which takes the form %
\begin{equation}
\hat{U}_{\mathrm{eff}}^{\left( 2D\right) }=\frac{U_{0}}{\left( 2\pi \right)
^{2}}\int_{\left\vert \vec{k}\right\vert ,\left\vert \vec{k}^{\prime
}\right\vert <k_{c}}|\vec{k})(\vec{k}^{\prime }|\otimes |\mathrm{S}\rangle
\langle \mathrm{S}|d\vec{k}d\vec{k}^{\prime },
\end{equation}%
with $k_{c}$ is a cut-off momentum. Similar as in the 3D
case, in the references where the area ${\cal S}$ of the system is first
assumed to be finite, the second-quantized form $\hat{U}_{\mathrm{eff}%
}^{\left( 2D\right) }$ is given by $\hat{U}_{\mathrm{eff}}^{\left(
2D\right) }=U_{0}^{\left( 2D\right) }{\cal S}^{-1}\sum_{\vec{k},\vec{k}^{\prime },%
\vec{K}}\,^{\prime }a_{\vec{K}/2+\vec{k},\uparrow }^{\dagger }a_{\vec{K}/2-%
\vec{k},\downarrow }^{\dagger }a_{\vec{K}/2-\vec{k}^{\prime },\downarrow }a_{%
\vec{K}/2+\vec{k}^{\prime },\uparrow }$ with $a_{\vec{p},\sigma
}^{\dagger }$\textbf{, }$a_{\vec{p},\sigma }$ and  $\sum_{\vec{k},%
\vec{k}^{\prime },\vec{K}}^{\prime }$ have similar definitions as
in Sec. IV.B. The renormalization relation for this contact potential can
be obtained from the condition that $\hat{U}_{\mathrm{eff}}^{\left(
2D\right) }$ and the realistic inter-atomic interaction should lead to the
same low-energy scattering amplitude. In the absence of SO-coupling, the
standard calculation gives~\cite{rr2d}%
\begin{equation}
\frac{1}{U_{0}^{\left( 2D\right) }}=-\frac{1}{(2\pi )^{2}}\int_{k^{\prime
\prime }<k_{c}}d\vec{k}^{\prime \prime }\frac{1}{\varepsilon +k^{\prime
\prime 2}},  \label{r2d}
\end{equation}%
where the physical parameter $\varepsilon >0$ is the binding energy of the
2D two-atom bound state for the cases without SO coupling.

In the presence of SO coupling, the effective scattering amplitude $f_{%
\mathrm{eff}}(t^{\prime }\leftarrow t)$ given by $\hat{U}_{\mathrm{eff}}$
can be obtained with the method used in Sec. IV. A. The straightforward
caclulation gives%
\begin{eqnarray}
&&f_{\mathrm{eff}}^{\left( 2D\right) }\left( t^{\prime }\longleftarrow
t\right)  \notag \\
&=&-\frac{4\pi ^{3}\langle \Psi _{t^{\prime }}^{\left( 0\right) }(0)|\mathrm{%
S}\rangle \langle \mathrm{S}|\Psi _{t}^{\left( 0\right) }(0)\rangle }{\frac{%
i\pi }{2}-\ln E_{t}^{1/2}+\frac{2\pi }{U_{0}^{\left( 2D\right) }}+\ln
k_{c}-2\pi \mathcal{F}^{(2D)}\left( E_{t}\right) }.  \notag \\
&&
\end{eqnarray}%
By requiring that $f_{\mathrm{eff}}^{\left( 2D\right) }(t^{\prime
}\leftarrow t)=f^{\left( 2D\right) }(t^{\prime }\leftarrow t)$ with $%
f^{\left( 2D\right) }(t^{\prime }\leftarrow t)$ given by Eq.~(\ref{f2d}), we
obtain the following renormalization relation for systems with SO coupling
in the limit $k_{c}\rightarrow \infty $:%
\begin{equation}
\frac{1}{U_{0}^{\left( 2D\right) }}=-\frac{1}{(2\pi )^{2}}\int_{k^{\prime
\prime }<k_{c}}d\vec{k}^{\prime \prime }\frac{1}{\varepsilon _{R}+k^{\prime
\prime 2}},  \label{r2de}
\end{equation}%
with $\varepsilon _{R}=4\exp (-2C)/d_{R}^{2}$. Comparing Eq.~(\ref{r2d})
with Eq.~(\ref{r2de}), we find that for the 2D contanct potential with
momentum cutoff, the form of the renormalization relation is not changed by
the SO coupling. In the presnece of SO coupling one only needs to replace $%
\varepsilon $ with $\varepsilon _{R}$.

\subsection{Huang-Yang pseudo-potential}

In the end of this section consider the Huang-Yang pseudo-potential~\cite%
{hyp} for the SO-coupled spin-$1/2$ fermi gas in a 3D system. In the $\vec{r}
$-representation, the Huang-Yang pseudo-potential~\cite{hyp} is given by
\begin{equation}
U_{HY}=4\pi a\delta \left( \vec{r}\right) \frac{\partial }{\partial r}\left(
r\cdot \right) ,
\end{equation}%
with $a$ the scattering length. Since $U_{HY}$ is a zero-range potential,
the scattering state with respect to $U_{HY}$, i.e., the solution of the
equation%
\begin{equation}
\left[ H_{0}+U_{HY}\right] |\psi _{t}^{(+)}(\vec{r})\rangle =E_{t}|\psi
_{t}^{(+)}(\vec{r})\rangle
\end{equation}%
with out-going boundary condition, should take the form
\begin{equation}
|\psi _{t}^{(+)}(\vec{r})\rangle =|\Psi _{t}^{(0)}(\vec{r})\rangle
\!+B_{t}G_{0}\left( E_{t};\vec{r},0\right) |{\mathrm{S}}\rangle \,
\end{equation}%
in the region with $r\equiv |\vec{r}|>0$, and thus satisfy the Bethe-Peierls
boundary condition. In the absence of SO-coupling, the Bethe-Peierls
boundary condition is%
\begin{equation}
\lim_{r\rightarrow 0}|\psi _{t}^{(+)}(\vec{r})\rangle \propto \left( \frac{1%
}{r}-\frac{1}{a}\right) |{\mathrm{S}}\rangle .
\end{equation}%
According to this condition, the function $r|\psi _{t}^{(+)}(\vec{r})\rangle
$ is continuous in the point $r=0$. Therefore, the partial derivative $%
\partial /\partial r$ in $U_{HY}$ is well-defined for the function $r|\psi
_{t}^{(+)}(\vec{r})\rangle $. Namely, the operation of the Huang-Yang
pseudo-potential $U_{HY}$ on the wave function $|\psi _{t}^{(+)}(\vec{r}%
)\rangle $ is well-defined.

Nevertheless, in the presence of the SO-coupling, as shown in Ref.~\cite%
{ourbp}, the Bethe-Peierls boundary condition is modified to
\begin{equation}
\lim_{r\rightarrow 0}|\psi _{t}^{(+)}(\vec{r})\rangle \propto \left( \frac{1%
}{r}-\frac{1}{a}\right) |{\mathrm{S}}\rangle -i\frac{\lambda }{2}\vec{c}%
\cdot \left( \frac{\vec{r}}{r}\right) |{\mathrm{S}}\rangle .
\end{equation}%
Due to the anisotropic term $-i\lambda \vec{c}\cdot \vec{r}/(2r)|{\mathrm{S}}%
\rangle $, the function $r|\psi _{t}^{(+)}(\vec{r})\rangle $ is \textit{not}
continuous in the point $r=0$. As a result, the partial derivative $\partial
/\partial r$ in $U_{HY}$ is not well-defined for the function $r|\psi
_{t}^{(+)}(\vec{r})\rangle $, and thus the operation of $U_{HY}$ on the wave
function $|\psi _{t}^{(+)}(\vec{r})\rangle $ is no longer well-defined.
Therefore, in the presentee of SO-coupling, the Huang-Yang pseudo-potential
is not consistent with the modified Bethe-Peierls boundary condition, and
thus cannot be directly used in the theoretical calculations.

\section{Discussion}

In this paper we derive the analytical expression of the scattering
amplitude of two ultra-cold atoms with SO coupling. Our approach can also be
used to calculate the wave function and the energy of two-body bound state.
Moreover, we show that the SO coupling can induce inter-atomic scattering
resonance. The influence of such a resonance in the many-body physics of the
SO-coupled ultra-cold gases remains to be explored. With the expression of
scattering amplitude, we further prove that the renormalization relations of
the 3D and 2D contact potentials with momentum cutoff are not changed by the
SO coupling. Nevertheless,  in the presence of SO coupling the
physical parameters in the renormalization relation should be replaced by
the ones which are related to the SO coupling. It is pointed out that,
our result provides a solid basis for some previous theoretical
works (e.g., Refs.~\cite{fewbody2,yu-11,zhang-12}) where  the
renormalizations relation with these forms are used  for the SO-coupled
gases without proof. We also show that in the presence of the SO-coupling,
the Huang-Yang pseudo-potential is no longer consistent with the modified
Bethe-Peierls boundary condition, and thus cannot be directly used in the
theoretical calculations.

\begin{acknowledgments}
We thank H. Zhai, X. Cui, H. Duan and L. You for useful discussions. This
work is supported by National Natural Science Foundation of China under
Grants No. 11074305, 11222430, 11275185, 10975127, CAS, NKBRSF of China
under Grants No. 2012CB922104, 2011CB921300. PZ would also like to thank
the NCET Program for support.
\end{acknowledgments}

\appendix
\addcontentsline{toc}{section}{Appendices}\markboth{APPENDICES}{}
\begin{subappendices}

\section{Behavior of scattering wave function}

In this Appendix we give the explicit definition of the effective range $%
r_{\ast }$ for the potential $U(\vec{r})$, and prove Eqs. (\ref{sw2}, \ref%
{btr}, \ref{21}, \ref{btr2}, \ref{psib}) for the behavior of the scattering
state $|\Psi _{t}^{(+)}(\vec{r})\rangle $ and bound state $|\Psi _{b}(\vec{r}%
)\rangle $ in the region $r\gtrsim r_{\ast }$. Although this problem has
been discussed by us in Refs.~\cite{ourbp} and~\cite{fewbody4}, the
treatment there is not fully rigorous. Here we provide a more explicit
analysis.

We first consider the case of spin$-1/2$ atoms. According to the scattering
theory, $|\Psi _{t}^{(+)}(\vec{r})\rangle $ is determined by the
Lippmman-Schwinger equation
\begin{equation}
|\Psi _{t}^{(+)}(\vec{r})\rangle =|\Psi _{t}^{(0)}(\vec{r})\rangle +\int d%
\vec{r}^{\,\prime }G_{0}(E_{t},\vec{r},\vec{r}^{\,\prime })U(\vec{r}%
^{\,\prime })|\Psi _{t}^{(+)}(\vec{r}^{\,\prime })\rangle ,  \label{bb1}
\end{equation}%
with the Green's function $G_{0}$ defined in Eq.~(\ref{g2d}). Here we assume
the potential $U(\vec{r})$ is negligible when the inter-atomic distance is
larger than a characteristic length $r_{0}$, i.e., $U(\vec{r})\simeq 0$ in
the region $r\gtrsim r_{0}$. Thus, the integration in Eq.~(\ref{bb1}) is
only effective in the region $r^{\prime }\lesssim r_{0}$.

If the momentum $k$ is low enough so that $k<<1/r_{0}$, when $r\rightarrow
\infty $ and $r^{\prime }\lesssim r_{0}$, the function $G_{0}(E,\vec{r},\vec{%
r}^{\,\prime })$ varies very slowly with respect to $\vec{r}^{\,\prime }$
and we have $G_{0}(E,\vec{r},\vec{r}^{\,\prime })\approx G_{0}(E,\vec{r},0)$%
. Therefore, in the limit $r\rightarrow \infty $, the solution of Eq.~(\ref%
{bb1}) takes the form
\begin{equation}
|\Psi _{t}^{(+)}(\vec{r})\rangle \approx |\Psi _{t}^{(0)}(\vec{r})\rangle
+G_{0}(E_{t},\vec{r},0)|\phi \rangle ,  \label{bb4}
\end{equation}%
where the spin state $|\phi \rangle $ is related to $|\Psi _{t}^{(+)}(\vec{r}%
)\rangle $ via the equation
\begin{equation}
|\phi \rangle =\int d\vec{r}^{\,\prime }U(\vec{r}^{\,\prime })|\Psi
_{t}^{(+)}(\vec{r}^{\,\prime })\rangle .  \label{bb8a}
\end{equation}%
Furthermore, due to the facts $\mathrm{P}_{12}|\Psi _{t}^{(+)}(-\vec{r}%
)\rangle =-|\Psi _{t}^{(+)}(\vec{r})\rangle $ and $\mathrm{P}_{12}U(-\vec{r})%
\mathrm{P}_{12}=U(\vec{r})$ with $\mathrm{P}_{12}$ the permutation operator
of the spin of the two atoms, one finds that $\mathrm{P}_{12}[U(-\vec{r}%
)|\Psi _{t}^{(+)}(-\vec{r})\rangle ]=-U(\vec{r})|\Psi _{t}^{(+)}(\vec{r}%
)\rangle $. This result yields
\begin{equation}
|\phi\rangle=\int d\vec{r}^{\,\prime }U(\vec{r}^{\,\prime })|\Psi _{t}^{(+)}(%
\vec{r}^{\,\prime })\rangle =B_{t}|\mathrm{S}\rangle\,,  \label{bb5}
\end{equation}%
with $B_{t}$ a constant number. Therefore, when the inter-atomic distance $r$
is \textit{large enough}, we have
\begin{equation}
|\Psi _{t}^{(+)}(\vec{r})\rangle \approx |\Psi _{t}^{(0)}(\vec{r})\rangle
+B_{t}G_{0}(E_{t},\vec{r},0)|\mathrm{S}\rangle .  \label{bbr}
\end{equation}

A direct result is that there exists a characteristic length $r_{\ast }$ and
Eq.~(\ref{bbr}) is applicable in the region $r\gtrsim r_{\ast }$. It is
apparent that we have $r_{\ast }\gtrsim r_{0}$. In this paper we define such
a characteristic length as the effective range of the potential $U(\vec{r})$%
. Therefore, Eqs.~(\ref{bbr}, \ref{bb8a}) or Eqs.~(\ref{sw2}, \ref{btr}) are
naturally satisfied when $r\gtrsim r_{\ast }$.

When $U(\vec{r})$ is spherical, in the low-energy case one only needs to
take into account the $s$-wave scattering. The straightforward calculaiton
shows that, as a result of the $s$-wave approximation, we have $r_{\ast
}=r_{0}$. When $U(\vec{r})$ is anisotropic, the effective range $r_{\ast }$
is larger than $r_{0}$. In our current paper and Ref.~\cite{ourbp}, we
assume the kinetic energy of atomic relative motion is low enough so that
the effective ranges of $U(\vec{r})$ and the \textquotedblleft rotated
potential"
\begin{eqnarray}
&&U_{\mathrm{R}}(\vec{r})\equiv  \notag \\
&&e^{i\lambda c_{z}x/2}e^{i\lambda c_{y}y/2}e^{i\lambda c_{x}z/2}U(\vec{r}%
)e^{-i\lambda c_{x}x/2}e^{-i\lambda c_{y}y/2}e^{-i\lambda c_{z}z/2}  \notag
\\
&&
\end{eqnarray}%
is much smaller than $1/k$. This condition can be satisfied in the dilute
gases. In particular, in the systems where the condition $r_{0}<<1/k$ is
satisfied and the anistropicity of $U(\vec{r})$ and $U_{\mathrm{R}}(\vec{r})$
is small enough so that the relevant effective ranges of $U$ and $U_{\mathrm{%
R}}$ are in the same order of magnitude with $r_{0}$.

Our above analysis can be directly generalized to the general cases of atoms
with arbitrary spin. In that case the scattering state $|\Psi _{t}^{(+)}(%
\vec{r})\rangle $ takes the form in Eq.~(\ref{bb4}) in the whole region with
$r\gtrsim r_{\ast }$, and the $\vec{r}$-independent state $|\phi \rangle $
is also related to $|\Psi _{t}^{(+)}(\vec{r})\rangle $ via Eq.~(\ref{bb8a}).
Nevertheless, now $|\phi \rangle $ is not unique. It can be different for
different incident states $|\Psi _{t}^{\left( 0\right) }(\vec{r})\rangle \!$%
. Then we can obtain the result in Eqs.~(\ref{21}, \ref{btr2}).

Now we consider the low-energy bound state of two SO-coupled atoms. It is
clear that the wave function $|\Psi _{b}(\vec{r})\rangle $ of the bound
state is given by the equation%
\begin{equation}
|\Psi _{b}(\vec{r})\rangle =N_{b}\int d\vec{r}^{\,\prime }G_{0}(E_{b},\vec{r}%
,\vec{r}^{\,\prime })U(\vec{r}^{\,\prime })|\Psi _{b}(\vec{r}^{\,\prime
})\rangle
\end{equation}%
with $E_{b}$ the energy of the bound state and $N_{b}$ the normalization
factor. With similar discussion as above, when $|E_{\mathrm{binding}%
}|<<1/r_{0}^{2}$ we have
\begin{equation}
|\Psi _{b}(\vec{r})\rangle \approx N_{b}G_{0}(E_{b},\vec{r},0)|\phi_b \rangle
\end{equation}%
in the region $r>r_{\ast }$. That is the result in Eq.~(\ref{psib}). In this
paper we assume the condition $|E_{\mathrm{binding}}|<<1/r_{\ast }^{2}$ is
satisfied.

\section{The short-range behaviors of free Green's function}

In this Appendix we prove Eq.~(\ref{gga}) for the short-range behaviors of $%
G_{0}(E_{t},\vec{r},0)$ with the following three steps.

First, with the fact
\begin{equation}
\delta \left( \vec{r}-\vec{r}^{\,\prime }\right) =\int d\vec{k}^{\prime
\prime }\frac{e^{i\vec{k}^{\prime \prime }\cdot \left( \vec{r}-\vec{r}%
^{\,\prime }\right) }}{\left( 2\pi \right) ^{3}}\left( \sum_{\alpha ^{\prime
\prime }}|\alpha ^{\prime \prime },\vec{k}^{\prime \prime }\rangle \langle
\alpha ^{\prime \prime },\vec{k}^{\prime \prime }|\right) ,
\end{equation}%
it is easy to show that
\begin{equation}
G_{0}(\eta ,\vec{r},0)=\sum_{\alpha }\int d\vec{k}^{\prime \prime }\frac{e^{i%
\vec{k}^{\prime \prime }\cdot \vec{r}}}{\left( 2\pi \right) ^{3}}\frac{%
|\alpha ^{\prime \prime },\vec{k}^{\prime \prime }\rangle \langle \alpha ,%
\vec{k}^{\prime \prime }|}{\eta +i0^{+}-E_{t^{\prime \prime }}}.
\label{bb10}
\end{equation}%
with $t^{\prime \prime }=(\alpha ^{\prime \prime },\vec{K},\vec{k}^{\prime
\prime })$. Eq.~(\ref{bb10}) and the completeness relationship $\sum_{\alpha
^{\prime \prime }}|\alpha ^{\prime \prime },\vec{k}^{\prime \prime }\rangle
\langle \alpha ^{\prime \prime },\vec{k}^{\prime \prime }|=1$ lead to the
result%
\begin{eqnarray}
&&G_{0}(\eta ,\vec{r},0)=\int d\vec{k}^{\prime \prime }\frac{e^{i\vec{k}%
^{\prime \prime }\cdot \vec{r}}}{\left( 2\pi \right) ^{3}}\frac{1}{\eta
+i0^{+}-k^{\prime \prime 2}}  \notag \\
&&+\int d\vec{k}^{\prime \prime }\frac{e^{i\vec{k}^{\prime \prime }\cdot
\vec{r}}}{\left( 2\pi \right) ^{3}}\mathcal{F}(\eta ,\vec{k}^{\prime \prime
}),  \label{bb12}
\end{eqnarray}%
with the operator $\mathcal{F}(\eta ,\vec{k}^{\prime \prime })$ defined in
Eq. (\ref{if}). Similarly as in Sec. II.A, the fact $\lim_{|\vec{k}%
|\rightarrow \infty }h_{0}(\vec{k})=\lambda \vec{c}{\cdot }\vec{k}$
gives $\lim_{k^{\prime \prime }\rightarrow \infty }[\mathcal{F}(\eta ,\vec{k%
}^{\prime \prime })e^{i\vec{k}^{\prime \prime }\cdot \vec{r}}+\mathcal{F}%
(\eta ,-\vec{k}^{\prime \prime })e^{-i\vec{k}^{\prime \prime }\cdot \vec{r}%
}]\propto \cos (\vec{k}^{\prime \prime }\cdot \vec{r})[1/k^{\prime \prime 4}+%
\mathcal{O}(1/k^{\prime \prime 5})]+\sin (\vec{k}^{\prime \prime }\cdot \vec{%
r})[1/k^{\prime \prime 3}+\mathcal{O}(1/k^{\prime \prime 4})]$.
Therefore, using $\int d\vec{k}^{\prime \prime }=\int_{0}^{\infty
}k^{\prime \prime 2}dk^{\prime \prime }\int d\Omega _{\vec{k}^{\prime \prime
}}$ with $\Omega _{\vec{k}^{\prime \prime }}$ the solid
angle of $\vec{k}^{\prime \prime }$ it is easy to prove that the
integration in the right-hand side of Eq.~(\ref{bb12}) does not diverge. On
the other hand, we also have%
\begin{equation}
\int d\vec{k}^{\prime \prime }\frac{e^{i\vec{k}^{\prime \prime }\cdot \vec{r}%
}}{\left( 2\pi \right) ^{3}}\frac{1}{\eta +i0^{+}-k^{\prime \prime 2}}=-%
\frac{e^{i\sqrt{\eta }r}}{4\pi r}.  \label{ba5}
\end{equation}%
with $\arg \eta \in (-\pi ,\pi ]$. Due to these facts, we have
\begin{equation}
\lim_{r\rightarrow 0}G_{0}(\eta ,\vec{r},0)=-\frac{1}{4\pi r}+\mathcal{O}%
(r^{0}).  \label{ba1}
\end{equation}

Second, as in Ref.~\cite{ourbp}, we introduce a unitary transformation $%
\mathcal{R}(\vec{r})$ as
\begin{equation}
\mathcal{R}({\vec{r}})=e^{i\lambda c_{z}x/2}e^{i\lambda c_{y}y/2}e^{i\lambda
c_{x}z/2},  \label{ug}
\end{equation}%
with $\vec{c}\equiv (c_{x},c_{y},c_{z})$. The rotated free Hamiltonian $H_{0%
\mathrm{R}}=\mathcal{R}(\vec{r})H_{0}\mathcal{R}^{\dagger }(\vec{r})$ can be
calculated as
\begin{equation}
H_{0\mathrm{R}}=\vec{p}^{\,2}-2\lambda \vec{d}(\lambda \vec{r})\cdot \vec{p}%
+W(\vec{r})  \label{hrr}
\end{equation}%
with operators $\vec{d}\equiv (d_{x},d_{y},d_{z})$ and $W$ given by
\begin{eqnarray}
d_{x}\left( \lambda \vec{r}\right) &=&0;  \label{d1} \\
d_{y}\left( \lambda \vec{r}\right) &=&e^{i\lambda c_{z}z/2}\frac{c_{y}}{2}%
e^{-i\lambda c_{z}z/2}-\mathcal{R}(\vec{r})\frac{c_{y}}{2}\mathcal{R}%
^{\dagger }(\vec{r}); \\
d_{z}\left( \lambda \vec{r}\right) &=&\frac{c_{z}}{2}-\mathcal{R}(\vec{r})%
\frac{c_{z}}{2}\mathcal{R}^{\dagger }(\vec{r}),  \label{d3}
\end{eqnarray}%
and%
\begin{eqnarray}
W(\vec{r}) &=&i\lambda \left[ \nabla \cdot \vec{d}(\lambda \vec{r})\right] +%
\mathcal{R}(\vec{r})B(\vec{K})\mathcal{R}^{\dagger }(\vec{r})+  \notag \\
&&\lambda ^{2}\left[ |\vec{d}(\lambda \vec{r})|^{2}-\mathcal{R}(\vec{r})%
\frac{|\vec{c}|^{2}}{4}\mathcal{R}^{\dagger }(\vec{r})\right] .  \label{w}
\end{eqnarray}

Now we define the rotated Green's function $G_{0\mathrm{R}}(\eta ,\vec{r},0)$
as the Green's function with respect to $H_{0\mathrm{R}}$, i.e.,%
\begin{equation}
G_{0\mathrm{R}}(\eta ,\vec{r},0)=\frac{1}{\eta +i0^{+}\!-\!H_{0\mathrm{R}}\!}%
\delta \left( \vec{r}\right) .
\end{equation}

According to Eqs.~(\ref{d1}-\ref{d3}), we have $\vec{d}(\lambda \vec{r})=%
\mathcal{O}(\lambda r)$. Namely, the SO-coupling term in $H_{\mathrm{R}}$
vanishes in the limit $r\rightarrow 0$. Due to this fact, in such a limit $%
G_{0\mathrm{R}}(\eta ,\vec{r},0)$ has the same behavior with the Green's
function $G_{W}(\eta ,\vec{r},0)$, which is defined as
\begin{eqnarray}
G_{W}(\eta ,\vec{r},0) &=&\frac{1}{\eta +i0^{+}\!-\left[ \vec{p}^{\,2}+W(0)%
\right] }\delta \left( \vec{r}\right)  \notag \\
&=&-\sum_{n}|W_{n}\rangle \langle W_{n}|\frac{e^{i\sqrt{\eta -W_{n}}r}}{4\pi
r},
\end{eqnarray}%
with $W_{n}$ and $|W_{n}\rangle $ the $n$-th eigen-value and eigen-state of
the operator $W(0)$, respectively. Since $\lim_{r\rightarrow 0}G_{W}(\eta ,%
\vec{r},0)$ is the sum of $-1/(4\pi r)$ and a $\vec{r}$-independent
operator, we have
\begin{equation}
\lim_{r\rightarrow 0}G_{0\mathrm{R}}(\eta ,\vec{r},0)=-\frac{1}{4\pi r}%
+G_{0}^{\prime }(\eta ),  \label{rab}
\end{equation}%
with $G_{0}^{\prime }(\eta )$ a $\vec{r}$-independent operator in the space
of two-atom spin.

As in Ref.~\cite{ourbp}, the result in Eq.~(\ref{rab}) can be proved as
follows. According to the definition of $G_{0\mathrm{R}}$, it is obvious
that
\begin{equation}
G_{0}(\eta ,\vec{r},0)=\mathcal{R}({\vec{r}})G_{0\mathrm{R}}(\eta ,\vec{r}%
,0).  \label{rel}
\end{equation}%
Due to this relation and Eq.~(\ref{ba1}), $G_{0\mathrm{R}}(\eta ,\vec{r},0)$
can be expressed as the sum of $-1/(4\pi r)$ and another term which
converges in the limit $r\rightarrow 0$. Then we have
\begin{eqnarray}
&&G_{0\mathrm{R}}(\eta ,\vec{r},0)=-\frac{1}{4\pi r}+\sum_{n=0}^{\infty
}r^{n}G_{n}^{\prime }(\eta )  \notag \\
&&+\sum_{l=1}^{\infty }\sum_{m_{l}=-l}^{l}\sum_{n=0}^{\infty
}r^{n}Y_{l,m_{l}}(\theta ,\phi )J_{l,m_{l},n}(\eta )\, ,  \label{aa2}
\end{eqnarray}%
with $\left( r,\theta ,\phi \right) $ the spherical coordinate. Here $%
Y_{l,m_{l}}(\theta ,\phi )$ are the spherical harmonic functions, $%
G_{n}^{\prime }(\eta )$ and $J_{l,m_{l},n}(\eta )$ are $\vec{r}$-independent
operators in the space of two-atom spin.

In the region $r>0$, the Green's function $G_{0\mathrm{R}}(\eta ,\vec{r},0)$
satisfies
\begin{eqnarray}  \label{aa1}
&&\left[ {\vec{p}}^{2}-2\lambda \vec{d}(\lambda \vec{r})\cdot \vec{p}+W(\vec{%
r})\right] G_{0\mathrm{R}}(\eta ,\vec{r},0) =\eta G_{0\mathrm{R}}(\eta ,\vec{%
r},0)\, .  \notag \\
\end{eqnarray}%
Substituting Eq.~(\ref{aa2}) into Eq.~(\ref{aa1}) and comparing the
coefficients of the term $r^{-2}$ in both sides, we find that because $%
d(\lambda \vec{r})=\mathcal{O}(\lambda r)$, one has $J_{l,m_{l},0}(\eta )=0.$
Therefore, in the limit $r\rightarrow 0$, $G_{0\mathrm{R}}(\eta ,\vec{r},0)$
behaves as in Eq.~(\ref{rab}).

Third, using Eq.~(\ref{rel}) and Eq.~(\ref{rab}), we finally have
\begin{equation}
\lim_{r\rightarrow 0}G_{0}(\eta ,\vec{r},0)=-\frac{1}{4\pi r}+i\frac{\lambda
}{8\pi }\vec{c}\cdot \left( \frac{\vec{r}}{r}\right) +G_{0}^{\prime }(\eta ).
\label{ba4}
\end{equation}%
Now we derive the $\vec{r}$-independent operator $G_{0}^{\prime }(\eta )$ in
the space of two-atom spin. Therefore, we have%
\begin{equation}
G_{0}^{\prime }(\eta )=\lim_{r\rightarrow 0}\frac{1}{2}\left[ G_{0}(\eta ,%
\vec{r},0)+G_{0}(\eta ,-\vec{r},0)+\frac{1}{2\pi r}\right] .
\end{equation}%
Using the Eqs. (\ref{bb12}) and (\ref{ba5}) and the fact%
\begin{eqnarray}
&&\lim_{r\rightarrow 0}\left[ \int d\vec{k}f(\vec{k})e^{i\vec{k}\cdot \vec{r}%
}+\int d\vec{k}f(\vec{k})e^{-i\vec{k}\cdot \vec{r}}\right]  \notag \\
&=&\lim_{r\rightarrow 0}\int d\vec{k}f(\vec{k})\left( e^{i\vec{k}\cdot \vec{r%
}}+e^{-i\vec{k}\cdot \vec{r}}\right)  \notag \\
&=&2\int d\vec{k}f(\vec{k}),
\end{eqnarray}%
we find that
\begin{equation}
G_{0}^{\prime }(\eta )=-\frac{i\eta ^{1/2}}{4\pi }+F\left( \eta \right) .
\label{sk}
\end{equation}%
with the operator $F\left( \eta \right) $ defined in Eq.~(\ref{ffeta}).

In the short-range region, the behavior of $G_{0}\left( \eta ;\vec{r}%
,0\right) $ is given by $\lim_{r\rightarrow 0}G_{0}(\eta ,\vec{r},0)$. Thus,
substituting Eq.~(\ref{sk}) into Eq.~(\ref{ba4}), we obtain
\begin{eqnarray}
G_{0}\left( \eta ;\vec{r},0\right) &\approx &-\frac{1}{4\pi }\left( \frac{1}{%
r}+i\eta ^{1/2}\right) +F\left( \eta \right) +i\frac{\lambda }{8\pi }\vec{c}%
\cdot \left( \frac{\vec{r}}{r}\right)  \notag \\
&&\ \ \ \ \ \ \ \ \ \ \ \ \ \ \ (\mbox{for }r_{\ast }\lesssim r<<1/k)\,.
\label{gg}
\end{eqnarray}%
That is the result in Eq.~(\ref{gga}).

\end{subappendices}

\end{document}